\input{aipcheck}

\documentclass[
    ,final            
,draft            
  ]
  {aipproc}

\layoutstyle{6x9}

\usepackage{amsmath,amssymb}

\begin{document}

\title{Quantum Geometry and Quantum Gravity}

\classification{04.60.Pp}
\keywords      {Quantum geometry, loop quantum gravity}

\author{J. Fernando Barbero G.}{
  address={Instituto de
Estructura de la Materia, CSIC, Serrano 123, 28006 Madrid, Spain}
}

\begin{abstract}
The purpose of this contribution is to give an introduction to quantum geometry and loop quantum gravity for a wide audience of both physicists and mathematicians. From a physical point of view the emphasis will be on conceptual issues concerning the relationship of the formalism with other more traditional approaches inspired in the treatment of the fundamental interactions in the standard model. Mathematically I will pay special attention to functional analytic issues, the construction of the relevant Hilbert spaces and the definition and properties of geometric operators: areas and volumes
\end{abstract}

\maketitle

\tableofcontents


\section{Introduction}

Loop quantum gravity (LQG) is one of the leading candidates to become the definitive quantum theory of general relativity. It is a very active field of research with potentially far reaching consequences as far as our ultimate understanding of Physics --and the universe itself-- is concerned. It is also a very rich and beautiful framework from the mathematical point of view. In fact the interplay between functional analysis and geometry that lies at the core of its mathematical fundations provides us with some very new ideas and a novel view about geometry that I will try to convey in these notes. I will start with a disclaimer: this is not intended to be a complete and up to date review of LQG. The reasons are manifold; lack of space is an obvious one, but also the availability of a number of excellent books and review articles that cover all the conceivable topics and points of view. I give some in the bibliography and refer to them those readers that want to delve into the depths of this subject. My intention here is to concentrate on two main topics with the purpose of motivating the readers to further study the beautiful problems of quantum gravity. One of them is the justification of the framework: why it makes sense to quantize gravity and, very specially, why it is so important to learn how to quantize diffeomorphism invariant field theories. In the following I have tried to give a balanced account of the reasons behind the failure of the traditional approaches to quantum gravity and the need to incorporate diffeomorphism invariance. I will also discuss what perturbative and non-perturbative approaches mean. The other subject is of a more mathematical nature: to introduce quantum geometry. Once we accept that gravity is geometry and that it must be quantized we must face the difficult question of understanding what geometry means in a quantum setting. The fact that this point can be satisfactorily dealt with is a triumph of LQG and, in my opinion, is a strong indication that we are in the right path to a definitive quantum description of gravity. Just to give an idea of the type of issues that I will not be touching here at all let me just mention a few such as the implementation of the quantum constraints, the building of the physical Hilbert space, loop quantum cosmology, spin foam models... As I just said before the reader is kindly directed to the vast literature available on all these subjects.

The paper is structured as follows. After this short introduction I will start discussing why it is necessary to quantize gravity, paying special attention to the role of diffeomorphisms and the difficulties that their correct quantum treatment entails. Afterwards I will give a quick introduction to the Hamiltonian formulation of gravity that is at the basis of LQG. To this end I will have to introduce the so-called Ashtekar variables that will allow us to describe gravity in a Yang-Mills phase space. The road to this formulation starts from the standard and well-known ADM framework and arrives at the new variables after enlarging the phase space and performing a suitable canonical transformation. I will briefly comment on the possibility of finding this Hamiltonian approach by using an action principle as the starting point.

I will devote the third section to discussing general aspects of the quantization with these variables. I will discuss in particular some differences between systems with a finite number of degrees of freedom and proper field theories, as far as the definition of the proper configuration space is concerned. This is easily done in the case of the scalar field so I will discuss this example here in some detail. Afterwards I will consider a similar construction for connection theories of the type considered here. I will give some details about the construction of the kinematical Hilbert space where the promised quantization of geometry is carried out paying special attention to the introduction of the relevant measure. I will also introduce a very useful orthonormal basis that will play an important role in the following: the spin network basis.

I quickly review in the fourth section some basic facts about the construction and main properties of geometric operators describing areas and volumes. Special attention will be paid to some important properties concerning their definition and their spectra that exhibit a telltale discretization of areas and volumes at a scale dictated by the Planck length. I finish the paper with an epilogue where I mention again some of the issues that have not been discussed here and ponder the future of the field and directions for future work.

\section{Why quantum gravity?}

Gravity was the first interaction known to mankind, though it took some time to be interpreted as such owing to its extreme feebleness. The other interaction traditionally known to us, electromagnetism (EM), is far stronger in real life. It is responsible for the cohesion of matter, the contact forces that keep things in place, friction... It also explains light and optics; in a very definite sense the modern technological world is directly shaped by our mastery and understanding of EM as formulated by Maxwell. The two remaining fundamental interactions: the so-called weak and strong forces are also very important (even technologically!) but their very short range means that their description must be intrinsically quantum. What does this mean?

Quantum mechanics of both finite and infinite-dimensional systems is a cornerstone of modern Physics. It is a necessary ingredient to understand the stability of matter and the intimate workings of nature; the kind of description that it provides has profound consequences as far as our understanding of the world is concerned. As I said before two of the fundamental interactions --the strong and weak nuclear forces-- are intrinsically quantum, their scales require a non-classical description right from the start. Even in the case of electromagnetism it is necessary to quantize it to save some potentially disastrous situations; in particular to explain the stability of matter. In all the three cases there is a successful framework, quantum field theory (QFT), that tells us how to correctly do this. QFT gives us the tools to understand how matter interacts and to pose the right experimental questions. Some important physical processes such as particle creation or nuclear reactions can only be understood within this framework. The so-called standard model of particles and interactions is a particular QFT.

Let me give here a very condensed summary of the key ideas and ingredients of the standard formulation of quantum field theories. To this end I will use as an example one of the simplest, yet impressively successful, models: Quantum electrodynamics (QED). A useful starting point is the Lagrangian
written in terms of a Dirac spinor $\psi$ and a 1-form EM field $A$.
$$
\mathcal{L}_{QED}=\bar{\psi}(i\partial\!\!\!\!\!/-m)\psi-\frac{1}{4}\mathrm{d}A\wedge\ast \mathrm{d}A+e\bar{\psi}A\!\!\!\!/\psi.
$$
As we can see it consists of two quadratic terms that define the \textit{free} part, and an extra \textit{interaction} term.  The setting in which the theory is formulated and a great deal of its physical interpretation is given by the free part of the Lagrangian, very specially when we consider its quantization. Quantizing means to build an appropriate Hilbert space of states and a representation of an algebra of elementary quantum operators representing the basic observables of the theory. In the case of field theories formulated in a fixed background --the Minkowski space-time of special relativity in this case-- it is also useful to represent the generators of the Poincar\'e group and some related objects (such as the Casimir operators) because they will allow us to give a physical interpretation to some vectors in the Hilbert space and label them in a convenient way.

The key idea to quantize the free part of the Lagrangian is to realize that being quadratic it can be interpreted as an infinite collection of independent harmonic oscillators. One can then use the well-known quantization of this basic system and adapt it directly by taking as the basic Hilbert space for the field theory an infinite tensor product of the one-oscillator Hilbert spaces. Though this is possible in principle (and, in fact, such a tensor product is well defined) there are some technical difficulties (separability, reducibility of the representation of the canonical commutation relations \cite{Wald 1}) that lead us to consider a different approach: using a so-called Fock space where these difficulties do not show up. The Hilbert space for a scalar field\footnote{A very similar construction gives the Fock space for fermion or vector fields of the types appearing in the QED Lagrangian written above. The main differences are the use of antisymmetrized tensor products to define the $n$-particle subspaces for fermion fields and the presence of labels for the spin degrees of freedom and the EM field.}, for example, has the form
$$\mathcal{F}_s(\mathcal{H})=\bigoplus_{n=0}^\infty\bigg(\otimes_s^n\mathcal{H}\bigg)=
\mathbf{C}\oplus\mathcal{H}\oplus\cdots;\quad\Psi=(\psi,\psi^{a_1},\psi^{a_1a_2},\ldots)\in \mathcal{F}_s(\mathcal{H}).$$

\noindent Its basic building block is the one-particle Hilbert space ${\mathcal{H}}$. The Fock space is an infinite direct sum of the so-called $n$-particle spaces given by the symmetrized products $\otimes_s^n\mathcal{H}$. At this point the reader should have guessed the physical interpretation of at least some of the vectors in the Hilbert space introduced above. To begin with the states in ${\mathcal{H}}$ are interpreted as elementary particles (electrons, positrons, and photons in the case of QED) with labels --masses and spin/helicities-- defined by the type of basic fields in the Lagrangian (fermion, vector, scalar). This is a consistent and fruitful interpretation because these vectors are eigenstates of the operators that represent some of the generators of the Poincar\'e group, in particular linear momentum and spin (helicity). They have the kind of properties that we associate with particles. In the somewhat picturesque language of QFT this is stated by saying that particles are the \textit{elementary quantum excitations} of the fields. This interpretation is further strengthened by the fact that the states in the $n$-particle subspaces can be obtained by the action of the so-called creation and annihilation operators on a special state $|0\rangle=(1,0,\cdots)$ called the vacuum (not to be confused with the zero vector!). A point that should be mentioned here is that some of the states in the Fock space describe a well defined number of elementary particles (and play a relevant role to describe the collision experiments carried out in particle accelerators) whereas other do not. An example of the latter are the coherent states that play a fundamental role in the physics of lasers and quantum optics. As a bona-fide Hilbert space $\mathcal{F}_s(\mathcal{H})$ is endowed with a scalar product that can be formally expressed as $\langle\Psi|\chi\rangle=\bar{\psi}\chi+\bar{\psi}_a\chi^a+\bar{\psi}_{a_1a_2}\chi^{a_1a_2}+\cdots$.

What about the other \textit{interaction} part of the Lagrangian? Important as the free part is to start interpreting the theory given by ${\mathcal{L}}_{QED}$, the model that a purely free Lagrangian describes is very boring from the physical point of view. Without a way to act on the system --interact with it-- we cannot do any Physics. In order to ``see'' a particle we must act upon it, or reciprocally, it should interact with something. Interaction implies a change in some of the properties --spins or momenta-- of the intervening elementary particles that are impossible to describe with a free Lagrangian. Here is where the other terms in the action come to the rescue. As we can see in ${\mathcal{L}}_{QED}$ these are simple monomials in terms of the basic fields. A very visual way to understand the role of these is to use the so-called Feynman diagrams as the one shown in figure 1.
\begin{figure}
  \includegraphics[width=.9\textwidth]{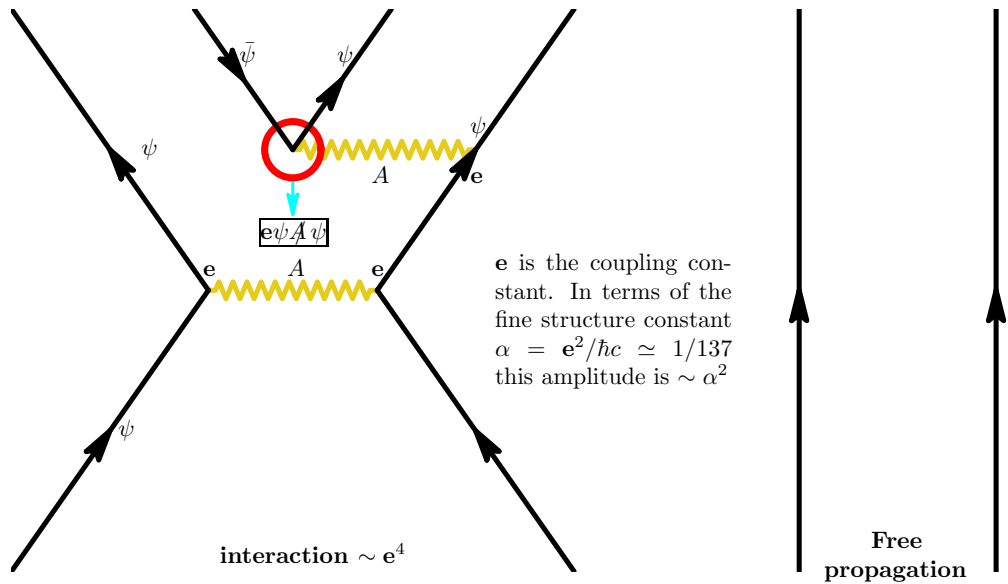}
  \caption{Feynman diagrams for a QED interaction process and pure propagation.}
\end{figure}
The diagram shown at the left of the picture describes the interaction of two electrons to give three electrons and a positron (or antielectron). This type of process, routinely detected in the laboratory these days, can only be naturally explained and described in the QFT framework that I am briefly discussing here.\footnote{The non-existence of a proper relativistic quantum mechnics is a direct consequence of the paradoxes posed by its inability to describe this type of process in a consistent way.} This is the right point to make several comments. The first one is related to how physical predictions are made. The object of direct experimental relevance is the so-called $S$-matrix that defines the cross sections that can measured in the lab. This cannot be computed in an analytic closed form and in practice is obtained as a power series in an interaction parameter. In the case of QED this is the famous fine structure constant $\alpha$. The fact that its value is small (approximately $1/137$ for low energies) means that successive amplitude terms added up to write an approximation of $S$ give smaller and smaller contributions to it (at least the first ones, as the power series expansion obtained in this way is known to be generically divergent!). Notice that the power of $\alpha$ that the amplitude described by a certain Feynman diagram gives is just half the number of the interaction vertices as shown in figure 1. A second important comment is related to the role played by the background Minkowskian metric that we use in the definition of the Lagrangian. Though non-dynamical it provides us with crucial structure for the quantum theory. The microcausality of the resulting quantum theory (that tells us that quantum observables at spatially separated space-time points are independent, i.e. commute) is dictated by it. Also the Poincar\'e group of symmetries of the Minkowskian metric not only tells us how the basic state vectors should be interpreted but also helps us select a unique quantization and avoid the huge indeterminacy otherwise present for the choice of a representation of the basic quantum observables \cite{Wald 1}.

What about gravity? Its ``true'' nature as spacetime geometry was recognized by Einstein in his attempts to find a relativistic description of gravitation (avoiding instant propagation and incorporating such pillars of relativity as the existence of a fundamental velocity).  A very important first step towards this was Minkowski's realization of the fact that special relativity could be understood as geometry in a spacetime endowed with a particular metric of signature $(-+++)$ --the Minkowski metric mentioned above. It should be clear now that at a very basic level the field theory that describes gravitation is rather different from the ones that describe the other basic interactions. In those, geometry plays an important but passive role, here it becomes a dynamical entity. How does one then quantize gravity? Can it be described as a QFT at a fundamental level?

These are rather old questions that have been addressed by theoretical physicists in many ways and for many years. There are two basic and quite different approaches to the subject that I will label here as the \textit{particle physicist} and the\textit{ relativist} points of view. To distinguish between them it is convenient to discuss how both of them answer the following set of basic questions: Should gravity be quantized? How do we do it? Can gravity be quantized?

Particle physics for the most part of the XX century has been a quest to understand the behavior of matter at higher and higher energies (though some significant low energy experiments such as the search for proton decay or other exotic processes have played a useful and complementary role). Particle theorists have strived, and succeeded, to predict and explain the experimental results at the highest energy regimes explored so far. The natural question would then be: What happens beyond those energies? Given that exploring higher energies also means probing shorter distances we expect that something significant should happen at the Planck scale where quantum mechanical and gravitational effects would both play a significant role\footnote{A hand-waving but quick and interesting way to see how this scale appears is to compute the energy $E$ of a photon having the property that the Schwarzschild radius corresponding to $E$ in a certain reference frame ($R_S=2G_N E/c^4$) is of the order of magnitude of its wavelength $\lambda=hc/E$. This gives $E\simeq(hc^5/G_N)^{1/2}$ i.e. the so-called Planck energy. Notice that this result can be arrived at by dimensional arguments just by combining the Planck constant $h$ relevant in quantum mechanics, the gravitational Newton constant $G_N$, and the speed of light $c$.}. Once a working description in these regimes is obtained one can use it to study the primitive universe and do cosmology in a more or less traditional way. As the understanding of higher energy behaviors has often led (or rather needed) a unified description of some of the fundamental interactions, there is a widespread belief within the particle physics community that a final consistent quantum description of gravity would require its unification with the other interactions. String theories are very important examples of this point of view.

Let me talk now about the particle physics attempts to quantize gravity within the so-called \textit{perturbative approach}. The natural goal is to obtain $S$ matrix elements containing the physics of particle interactions, giving predictions for accelerator experiments, and providing necessary information for cosmology. The use of perturbation theory for this means that one wants to obtain cross sections as power series in $G_N$ (as one does in QED). There are some technical complications that appear when one tries to compute amplitudes with higher order terms in the coupling constant in generic interacting QFT's. These show up whenever closed loops appear in the relevant Feynman diagrams. The reason is that the resulting amplitudes are infinite (because they involve the computation of divergent momentum integrals). In the case of the standard model there is a consistent (though arguably ugly) way to keep the infinite terms under control and obtain meaningful (in fact, fantastically accurate!) physical predictions. This is the reason why these theories are called renormalizable. Renormalizability is a key technical condition that QFT's must satisfy in order to have predictive power. What happens then in the case of gravity? Is it a renormalizable theory too? The answer, known since the late seventies and early eighties (after the pioneering works of 't Hooft, Veltman, Goroff, and Sagnotti) is negative. So, at least from the standard approach to the problem,  the question as to whether gravity can be quantized has a negative answer for a particle physicist. It must be said, however, that string theories and their derivatives may well provide a different answer because in some sense they are capable of incorporating gravity in a consistent way (infinites of the type present in the ordinary perturbative approach appear to be absent in string theories). In my opinion it is fair to say that there is not a final word on this issue.

Let us discuss now the other alternative point of view by asking the following question: what kind of phenomena would a relativist like to explore if a consistent theory of quantum gravity were available? Probably the first point would be to understand singularities because, in a certain sense, quantum field theories help solve similar problems in classical electrodynamics. Related to this would be the general problem of understanding the high energy regime of gravity in a cosmological setting (the beginning of the universe and the Big Bang) or in astrophysical situations (black holes). There is a widespread perception that the appearance of singularities in a physical theory is a clear indication that there is a better one (that reduces to the old one in a certain low energy, large distance scale but supersedes it outside these regimes) where these are not present. Other relevant questions from a relativist's point of view are related to black hole evaporation, the meaning of black hole entropy and issues related to information loss and the unitarity of quantum evolution. A rather characteristic trademark of some of the different approaches to the quantization of gravity on this side is related to their mathematical formulation. In fact it is rather significant, in my opinion, that some of the leading approaches, and very specially loop quantum gravity, pay special attention to mathematical rigor and consistency. This is perceived as a necessity given the impossibility of having an experimental (or observational) guide of the type that so successfully led to the development of the standard model. A short list of other interesting questions relevant here would include the problem of time and space-time covariance in general, the meaning of quantized geometry, and the emerging of classical geometry from a quantum gravity theory.

The main techniques that relativists have used to attempt a quantization of gravity have relied on canonical methods, and in particular Dirac's approach to the treatment of constrained systems. The starting point here is a Hamiltonian formulation for gravity. This was first obtained in the early sixties by Arnowitt, Deser, and Misner (the ADM formalism \cite{ADM}) after some pioneering work by Dirac. The first attempts to write a quantum version of the constraints led to significant insights into the possible features of quantized gravity but in the end failed to produce a consistent theory. Some of the issues that proved very difficult to solve within this approach were of a mathematical nature and posed difficult obstacles comparable to the lack of renormalizability of the perturbative approaches. There have been exciting developments in recent years that give us the hope that canonical quantization may, in the end, be the way to get to quantum gravity. Anyway, let us postpone its discussion for a while and stop to discuss an issue that will prove central in the following: the role of diffeomorphism invariance.

\subsubsection{The case for diffeomorphism invariance}

Let me start now by very quickly reviewing the so-called \textit{background field method} in standard QFT. This is an important and seldom discussed issue that is necessary to tackle in order to understand some of the problems posed by the quantization of general relativity. This method appeared as an efficient way to compute gauge invariant counterterms in Yang-Mills theories (with the help of the so-called covariant gauges, see for example \cite{Wein}). The starting point is a simple splitting of the basic fields $\phi$ appearing in a certain Lagrangian as $\phi=\phi_{back}+\varphi$ where $\phi_{back}$ is a \textit{fixed}, non-dynamical object (in the sense that there are no field equations associated to variations $\delta\phi_{back}$) and $\varphi$ is a new dynamical field (sometimes referred to, quite whimsically, as a ``quantum perturbation''). From the point of view of the path integrals profusely used in the usual approaches to QFT this is just a simple change of variables (in fact an ``innocuous translation'' of the integration variables). Notice that although $\varphi$ is usually called a ``quantum perturbation'' there is no reason to consider it small in any sense, at least in the framework of Yang-Mills theories, because the Lagrangian is polynomial and no convergence issues arise. Notice also that it is not strictly necessary to use this method; one can also work with the original variables and get equivalent results in the computation of counterterms. The advantage of using the background field method is a purely techical one: computations are much easier to perform in this scheme.

Let us discuss now what happens if one tries to use the same approach to quantize gravity with the techniques of standard, particle physics oriented, QFT. The first difficulty that we must face is the fact that the Einstein-Hilbert action is not polynomial, there is no natural splitting as a kinetic term plus interactions. As I discussed before the identification of a suitable free part in the Lagrangian is a fundamental first step because it determines the Hilbert space where the quantization will be carried out. This will tell us, among other things, what the fundamental quantum excitations are, their physical interpretation and properties\footnote{If the reader is interested in knowing what diff-invariant free actions look like and the kind of physics that they may describe see \cite{Barbero:1999ts,BarberoG.:2002fd}.} Here the background field method comes to the rescue because by writing the space-time metric as $g_{ab}=g_{ab}^{back}+h_{ab}$ and formally expanding the Einstein-Hilbert Lagrangian as a power series in $h_{ab}$ we can actually obtain an action with (more or less) the desired form. Specifically we find a free part (kinetic term) as in the familiar Yang-Mills case, plus an infinite collection of interaction terms. If $g_{ab}^{back}$ is taken to be the Minkowski metric $\eta_{ab}$ the free part describes the so-called Fierz-Pauli Lagrangian and the corresponding Fock Hilbert space describes massless, helicity two, particle-like excitations known as gravitons. Notice that it is precisely the Poincar\'e invariance introduced with the Minkowski metric that allows us to talk about particle properties such as mass, spin, helicity, and even energy and momenta. It is very important to realize that the description of gravity in terms of gravitons is a direct consequence of both our choice of the background field method and the specific splitting of the metric as $g_{ab}=\eta_{ab}+h_{ab}$. In this setting, where the action is not polynomial, convergence issues may play a role. The hope is of course that for ``small enough'' quantum perturbations $h_{ab}$, a power series expansion such as the one discussed above is well defined. This is the reason why this point of view is sometimes referred to as ``perturbative quantum gravity''. Notice that ordinary field theories (QED for example) are not perturbative in this sense but rather due to the fact that physically interesting quantities can only be obtained as power series in the relevant coupling constants (of course, this would also be the case for the perturbative quantum gravity that I am discussing here). The Minkowskian metric introduced in the previous splitting is also important because it provides the causal structure which plays a crucial role in standard approaches to quantum field theory (microcausality, spin-statistics connection,...).

The use of the background field method in quantum gravity is very problematic. Of course a first and obvious problem is that perturbative quantum gravity is a bad theory because it is not renormalizable. In practice this means that it is not possible to make physical sense of it. Other problems are more philosophical in nature: What is the right background? From the particle physics point of view where experiments are local in nature, Minkowski is a natural choice, but what if we are interested in the universe as a whole? Notice that topological issues would become important and may preclude the use of a Minkowskian background. Finally there is a clear conflict with some of the things that we would like to understand if a quantum theory of gravity were available: Does it make sense to use singular backgrounds? If not, then how would the quantized theory describe the very high curvature regimes expected in the vicinity of classical singularities?

An often mentioned difficulty with this type of approach is the fact that the introduction of a fixed non-dynamical object such as the Minkowski metric $\eta_{ab}$ breaks the diff-invariance of the theory. Some comments concerning this point are in order now. First of all it is indeed true that diff-invariance is broken but this happens in a rather innocent way. The set of field equations in terms of $\eta_{ab}$ and $h_{ab}$ is completely equivalent to the original Einstein equations and, in fact, they have a set of symmetries that is directly related with those of the original equations. The problem, in my opinion, lies in the lack of any physically preferred background. In fact it is very natural to attempt a quantization \textit{without the use} of a background. It is at this point that diff-invariance enters the game in a crucial way because we have then to face the quantization of a theory defined by the diff-invariant Einstein-Hilbert action. As a side comment it is probably relevant to point out here that diff-invariance has some unexpected consequences for the perturbative formulation of quantum field theories. As shown in \cite{Barbero:1999ts,BarberoG.:2002fd} it is impossible to build a quadratic diff-invariant action with local degrees of freedom. This implies that it is not possible to define a concept of elementary particle at a diff-invariant level. This strongly suggests that the elementary quantum excitations of the gravitational field must have an unusual nature. How this is realized in concrete terms within the framework of loop quantum gravity is one of the main topics in this paper.

A surprising development in this state of affairs was a key insight by Abhay Ashtekar \cite{Ashtekar:1986yd,Ashtekar:1987} who realized that gravity could be thought of as a theory of $SU(2)$ connections. This means, in particular, that it can be described in the same kind of phase space used in the canonical formulation of Yang-Mills theories. This has the desirable consequence that some of the methods already developed in this framework, and a lot of the associated geometrical machinery, can be used to advance in the hard problem of quantizing gravity. This led to a non-perturbative canonical gravity program that is widely known as \textit{loop quantum gravity} (LQG) these days. A word of caution concerning the meaning of non-perturbative is in order here. As I said before, there are very good reasons to avoid the introduction of background metrics in quantum gravity. This is certainly done in the Ashtekar variables approach because, in fact, metrics play a secondary role: The basic configuration variable is a gauge connection. It is in this sense that this approach to quantization is non-perturbative. On the other hand it may be unavoidable to develop some kind of approximation scheme to obtain sensible physical predictions in a usable form once a consistent quantum gravity theory is developed along these lines. In fact it is not realistic, in my opinion, to expect that the full theory will allow us to obtain an exact description of quantum states (and their scalar products) in the Physical Hilbert space when the classical limit --general relativity-- is such a complex physical theory.

\subsubsection{A prelude to quantization: Hamiltonians for general relativity}

The starting point for the quantization of a physical system is describing it in a Hamiltonian framework. This leads us to ask ourselves if General Relativity can be written in Hamiltonian form. The answer has been known since the late fifties and early sixties after some pioneering work by Dirac and Arnowitt, Deser and Misner \cite{ADM}.

A standard way to obtain a Hamiltonian description for a physical system is to derive it from an action principle. In the case of general relativity a good starting point is the Einstein-Hilbert action
$$
S=\frac{1}{2\kappa}\int_{\mathcal{M}}\mathbf{e}\sqrt{\sigma g}R;\quad \kappa=\frac{8\pi G_N}{c^3}.
$$
Here $(\mathcal{M},g_{ab})$ is a globally hyperbolic 4-dimensional manifold $\mathcal{M}=\mathbf{R}\times\Sigma$ where $\Sigma$ is chosen to be a smooth, orientable, closed (i.e. compact and without boundary) manifold, and $\sigma$ denotes the space-time signature ($\sigma=-1$ for Lorentzian space-times and  $\sigma=+1$ for the Riemannian ones). $R$ denotes the scalar curvature of a metric $g_{ab}$ and $\mathbf{e}$ a fiducial volume form. In order to arrive at a Hamiltonian description it is necessary to foliate the space-time manifold. This is achieved by means of a scalar ``time function" $t$ whose level hypersurfaces $\Sigma_t$ are smooth submanifolds diffeomorphic to  $\Sigma$. We have to introduce also a congruence of smooth curves nowhere tangent to these spatial hypersurfaces and parameterized by $t$. These define a time flow direction (i.e. a globally defined smooth vector field $t^a$ such that $t^a\nabla_a t=1$). A comment on notation: here and in the following I will use the abstract Penrose notation so indices do not refer to particular coordinate charts but are used as labels to identify the type of tensors that we are working with. A given metric $g_{ab}$ in ${\mathcal{M}}$ can be described in terms of three objects defined in terms of the previous foliation and congruence of curves: the lapse $N$, the shift $N^a$ and the 3-dimensional metric $h_{ab}:=g_{ab}-\sigma n_an_b$ defined with the help of the time-like, future-directed, unit normal to $\Sigma_t$. In order to arrive at a (3+1)-dimensional description we also need to introduce the unique, torsion-free, derivative operator $D_a$ on each $\Sigma_t$, compatible with the induced metric $h_{ab}$, and the extrinsic curvature $K_{ab}:=h_a^{\,\,c}\nabla_c n_b=\frac{1}{2N}({\mathcal{L}}_th_{ab}-2D_{(a}N_{b)})$. It is important to realize at this point the the information contained in $g_{ab}$ is the same as the one present in $(N,N^a,h_{ab})$ but these are ``3-dimensional objects'' in a definite sense, i.e. tensor fields defined on each of the $\Sigma_t$, and will then be useful in the (3+1)-dimensional description that we are searching for. The Einstein-Hilbert action can be rewritten in terms of these objects (it is useful to introduce a fiducial, non-dynamical volume form $\mathbf{e}:=e_{abcd}$ satisfying $\mathcal{L}_te_{abcd}=0$ as in \cite{Waldbook} and $^{(3)}e_{abc}:=t^de_{dabc}$). We get\footnote{From now on we use units such that $16\pi G_N/c^3=1$}
$$S=\int_{\mathcal{M}} \mathbf{e} \sqrt{h}N[\sigma{^{(3)}}R+K_{ab}K^{ab}-K^2]:=\int_{\mathbf{R}}\int_{\Sigma_t} {^{(3)}}\mathbf{e} \mathcal{L}_G:=\int_{\mathbf{R}}L_G.$$
As we can see the action is a time integral of a Lagrangian defined on the spatial manifold $\Sigma$. The integrand is written in terms of $t$-dependent tensor fields on $\Sigma$ and their time derivatives (remember that the extrinsic curvature $K_{ab}$ can be written as a time derivative of $h_{ab}$ and several factors involving the lapse and shift). The configuration space is given by the lapse, shift and 3-metrics.

At this point it is straightforward to arrive at a Hamiltonian formulation. We need only to introduce the canonically conjugate momenta to the configuration variables used above $p^{ab}:=\frac{\partial\mathcal{L}_G}{\partial \dot{h}_{ab}}=\sqrt{h}(K^{ab}-Kh^{ab})$ by performing a Legendre transform.
$$S=\int_{\mathbf{R}}\int_{\Sigma_t} {^{(3)}}\mathbf{e} \mathcal{H}_G:=\int_{\mathbf{R}}H_G,$$
$$\mathcal{H}_G=p^{ab}\dot{h}_{ab}-\mathcal{L}_G=N\bigg[\sigma\sqrt{h}^{(3)}R+\frac{1}{\sqrt{h}}(p^{ab}p_{ab}-
\frac{1}{2}p^2)\bigg]-2N_bD_a(h^{-1/2}p^{ab}).$$
Several comments are in order at this point. First of all we notice that no time derivatives of the lapse and shift appear in the Hamiltonian. This means that they behave as Lagrange multipliers enforcing the constraints
$$
\displaystyle\sigma\sqrt{h}\,^{\scriptscriptstyle(3)}\!R+\frac{1}{\sqrt{h}}(p^{ab}p_{ab}-
\frac{1}{2}p^2)=0,$$
$$
D_a(h^{-1/2}p^{ab})=0
$$
known as the scalar (or Hamiltonian) and vector constraint respectively. Second, the evolution of $h_{ab}$ and $p^{ab}$ are obtained directly from the Hamilton equations
$$
\dot{h}_{ab}=\frac{\delta H_G}{\delta p^{ab}},\hspace{1cm} \dot{p}^{ab}=-\frac{\delta H_G}{\delta h_{ab}}
$$
or, in terms of the Poisson brackets defined by the symplectic form given by the previous action
$$
\dot{h}_{ab}=\{h_{ab},H_G\},\quad \dot{p}^{ab}=\{p^{ab},H_G\}.
$$
Finally notice that in the compact case that we are considering here the only terms appearing in the Hamiltonian correspond to constraints; there are no extra terms.\footnote{These would be present, however, in the presence of spatial boundaries or in the non-compact case.}
The final Hamiltonian description that we arrive at is formulated in a phase space coordinatized by the spatial metrics $h_{ab}$ and their canonically conjugate momenta $p^{ab}$ (the symplectic structure is then the canonical one). The full dynamics of the system is given by the first class constraints written above. They generate gauge transformations. The ones generated by the vector constraint are easy to describe: they correspond to diffeomorphisms on the spatial slices $\Sigma$ whereas the remaining ones are much harder to deal with because, in a definite sense, they contain all the non-trivial dynamics of general relativity. In fact it is fair to say that at the present stage the main difficulties separating us from a completely satisfactory theory of quantum gravity are related in one or another form to the treatment of the Hamiltonian constraint.

The passage to the Ashtekar description of general relativity \cite{Ashtekar:1986yd,Ashtekar:1987} can be done in two steps. The first one
one is straightforward and consists in the introduction of an additional internal $SO(3)$ symmetry in the system. At this stage this is a little more than a simple change of variables. We start by introducing a triad i.e. three 1-form fields $e_a^i$, $i=1,2,3$ on $\Sigma$ defining a local co-frame at every point ($\det e\neq0$). In terms of them the 3-metric can be written as $h_{ab}=e_a^ie_b^j\delta_{ij}$. With the help of the  inverse triad $e^a_i$ we can build the ``densitized inverse triad'' $\tilde{E}^a_i=(\det e)e^a_i$. We can also define an object closely related to the extrinsic curvature $K_a^i:=K_{ab}\tilde{E}^b_j\delta^{ij}/\det e$. It is straightforward at this point to check that these new variables can be taken to be canonical (in spite of the fact that the previous construction is not a canonical transformation because the number of new variables is different from the number of old ones). The constraints in the usual ADM formulation can be written in terms of these new variables  but is important to realize that now we have local $SO(3)$ rotations of $e_a^i$ and $K_a^i$ so there must be extra constraints to generate them. These can easily be found from the condition that the extrinsic curvature is a symmetric tensor , i.e. $K_{[ab]}=0$. We arrive then at a phase space description where the canonical variables are now the pairs $(K_{a}^i,\tilde{E}^{a}_i)$ and the first class constraints become
\begin{eqnarray}
& & \epsilon_{ijk}K_{a}^{j}\tilde{E}^{ak}=0,\nonumber\\
& & {\cal D}_{a}\left[\tilde{E}^{a}_{k}K_{b}^{k}-\delta_{b}^{a}
\tilde{E}^{c}_{k}
K_{c}^{k}\right]=0,\label{1}\\
& & -\sigma\sqrt{h}R+\frac{2}{\sqrt{h}}
\tilde{E}^{[c}_{k}\tilde{E}^{d]}_{l}K_{c}^{k}K_{d}^{l}=0\nonumber
\end{eqnarray}
where $R$ denotes now the scalar curvature of $h_{ab}:=e_a^ie_{bi}$ and ${\cal D}_{a}$ is defined in such a way that ${\cal D}_{a}\tilde{E}^{a}_{i}=0$. As the reader can see the formulation is very similar to the standard ADM approach. However this changes in a very important way if one performs the following transformation
\begin{equation}
 _{\gamma}\tilde{E}^{a}_{i}= -\frac{1}{\gamma}\tilde{E}^{a}_{i},\quad _{\gamma}A_{a}^{i}=\Gamma_{a}^{i}+\gamma K_{a}^{i}\nonumber
\end{equation}
where the Immirzi parameter $\gamma$ \cite{Barbero:1994ap,Immirzi} is, at this stage, just a non-zero complex number and $\Gamma_a^i$ is the $SO(3)$ connection defined by the densitized inverse triad $\tilde{E}^{a}_{k}$ from the condition
$\partial_{[a}e_{b]}^i+\epsilon^{i}_{\,\,jk}\Gamma^j_{[a}e_{b]}^k=0.$
Notice that the new variable $_{\gamma}A_{a}^{i}$ is a $SO(3)$ connection. It is obvious now that  $\{_{\gamma}A_{a}^{i}(x),\,
_{\gamma}\tilde{E}^{b}_{j}(y)\}=\delta_{j}^{i}\delta_{a}^{b}\delta^{3}(x,y)$ and $\{_{\gamma}\tilde{E}^{a}_{i}(x),\,  _{\gamma}\tilde{E}^{b}_{j}(y)\}=0$. A slightly less trivial computation gives also $\{_{\gamma}A_{a}^{i}(x), \,_{\gamma}A_{b}^{j}(y)\}=0$ and, hence, we conclude that the previous transformation is, in fact, canonical. We can then describe gravity in a Yang-Mills phase space where the configuration variable has the interpretation of a $SO(3)$ connection and its canonical momentum is a ``$SO(3)$ electric field''. The constraints are now
\begin{eqnarray}
& & \nabla_a{_{\gamma}}\tilde{E}^{a}_i=0,\nonumber\\
& & F_{ab}^i{_{\gamma}}\tilde{E}^{b}_i=0,\nonumber\\
& & _{\gamma}\tilde{E}^{[a}_i {_{\gamma}}\tilde{E}^{b]}_j\left[\epsilon_{ijk}F_{ab}^k+
\frac{2(\sigma-\gamma^2)}{\gamma^2}({_{\gamma}}A_a^i-\Gamma_a^i)({_{\gamma}}A_a^i-\Gamma_b^j)\right]=0,
\nonumber
\end{eqnarray}
where $\displaystyle\nabla_a{_{\gamma}}\tilde{E}^{a}_i:=\partial_a{_{\gamma}}\tilde{E}^{a}_i
    +\epsilon_{ijk}{_{\gamma}}A_{a}^j{_{\gamma}}\tilde{E}^{ak}$, and $F_{ab}^i=2\partial_{[a}{_{\gamma}}A_{b]}^i+\epsilon^{ijk}{_{\gamma}}A_{aj}{_{\gamma}}A_{bk}$ is the $SO(3)$ curvature.
Some comments are necessary at this point. The first is that general relativity is formulated now in a Yang-Mills phase space. The fact that the configuration variable is a $SO(3)$ connection is, in fact, a cornerstone of the formalism and plays a crucial role in the quantization of the theory as discussed in the following. An interesting point that merits some attention is the role played by the Immirzi parameter $\gamma$. If the space-time signature is Riemannian we can cancel the last term in the Hamiltonian constraint by choosing $\gamma=\pm 1$. By doing this we get a very simple Hamiltonian constraint and the basic canonical variables are real. In the Lorentzian signature case the situation is not as straightforward as before. In fact we must choose between two possibilities: either we cancel the second (ugly) term in the Hamiltonian constraint by taking $\gamma=\pm i$ (in which case we have the extra problem that the canonical variables become complex) or we choose a real value for $\gamma$
and keep the somewhat complicated second term. For some time there was the widespread belief that it was better to work with the simplest Hamiltonian constraint and try to use some reality conditions to recover the physical real formulation. It was later realized by Thiemann that the second complicated term in the scalar constraint could actually be written in a manageable way with the help of some geometric operators and, thus, it was possible to make sense of it after quantization. This parameter shows up in the definition of area and volume operators and, as a consequence, in some physically important physical magnitudes such as black hole entropy. A somewhat surprising result is that a single choice of $\gamma$ suffices to get the right entropy (i.e. the one given by the Bekenstein formula) for all types of physically realistic black holes. It also plays a significant role in loop quantum cosmology.

Let me finish this section by pointing out that the previous Hamiltonian formulation can be derived from the following action principle \cite{Holst}
$$S=\int_{\mathcal{M}}e^I\wedge e^J\wedge(\epsilon_{IJKL}\Omega^{KL}-\frac{2}{\gamma}\Omega_{IJ}).$$
where $e^I$ is defined on a 4-dimensional $\mathbf{R}$-vector space ($I=0,1,2,3$), $\omega^I_{\,\,J}$ takes values in the Lie algebra of $SO(1,3)$ (for Lorentzian signature) or $SO(4)$ (for Riemannian signatures), $\epsilon_{IJKL}$ is the alternating tensor in $V$, and $\Omega^I_{\,\,J}=\mathrm{d}\omega^I_{\,\,J}+\omega^I_{\,\,K}\wedge\omega^K_{\,\,J}$ is the curvature 2-form of $\omega^I_{\,\,J}$. If $\gamma=i$ (Lorentzian case) or $\gamma=1$ (Riemannian case) the action can be written in terms of the self-dual curvature of a self-dual connection $\omega^{+I}_{\,\,\,\,J}$. There are two important comments about this. The first one is that this action is invariant under diffeomorphisms of $\mathcal{M}$ and ``internal'' gauge transformations ($SO(1,3)$ or $SO(4)$ depending on the signature). In order to arrive at the standard phase space formulation it is necessary to perform a partial gauge fixing (so that the only internal symmetry left is the $SO(3)$ symmetry introduced in the Hamiltonian framework). A second comment has to do with the structure of the action itself. As can be seen the first term alone suffices to obtain the Einstein equations. The second term does not change them but it is not a total divergence so the Holst action is not the same as the (Hilbert-Palatini) standard one. This is a non-trivial example where the possibility of obtaining the same field equations from two different actions plays a significant role.

\section{Quantization in terms of the new variables}

This section is devoted to the quantization of general relativity in terms of Ashtekar variables. A first comment that must be made before proceeding further is that our Hamiltonian formulation is, in a definite sense, different from the one for standard quantum mechanical systems (where we have a Hamiltonian and no constraints). Here we have constraints but we do not have a Hamiltonian. The scheme that we will use here is the so-called Dirac quantization for constrained systems. The main steps are the following: start by choosing a Poisson $*$-algebra of elementary classical variables (a family of functions in phase space that separate points; in classical mechanics one usually takes $q$ and $p$). Obtain a representation of this algebra in a suitable \textit{kinematical} Hilbert space $\mathcal{H}_{kin}$ of ``quantum states''. This will specifically force us to look for an appropriate linear space capable of accommodating the vectors describing our system. We have to choose also an inner product (this will lead us to ask ourselves for the relevant measures), and, finally, find physically interesting orthonormal bases where some important operators take simple forms (are diagonal, for example). Once the representation of the basic quantum algebra is found we have to obtain self-adjoint operators representing the constraints. This must be defined on $\mathcal{H}_{kin}$ or in the dual of a certain dense subspace of it. Alternatively one can represent elements of the group of symmetry transformations generated by the constraints. By looking for vectors in the kernels of these operators (and a suitable \textit{physical} inner product) one obtains physical states. If this process can be completed we still have to face the very important problem of extracting physics. This will specifically require that we find a set of self-adjoint operators in the physical Hilbert space, commuting with all the constraints, representing observables. Among them one should select the best suited for measurements. Concrete predictions with observable (astrophysics, cosmology) or experimental consequences must be made! Another important issue that must be tackled at this stage is to understand the classical limit (i.e. find out how to recover the macroscopic space-time geometry from the quantum model). This is straightforward for free theories (for example pure electromagnetism) but highly non-trivial in the presence of interactions. Notice that even for simple quantum mechanical systems such as the hydrogen atom this is highly non-trivial (in fact, no coherent states are known in this case). As I mentioned in the previous section it will be necessary, almost certainly,  to develop some sort of approximation scheme (a new perturbation theory) to get physical predictions.

Let us start then by constructing the kinematical Hilbert space. It is interesting at this point to consider a simple example that will help us put things in perspective: the free scalar field in Minkowski space-time. This is very useful because the quantization of this system is well understood from several points of view (remember that we can quantize it in a Fock space as mentioned in the introduction). A concept that plays an important role here is that of \textit{quantum configuration space}. On the one hand this will help us to get a description of the Hilbert space for a field theory that is somewhat related to the standard view of quantum mechanical models with a finite number of physical degrees of freedom, on the other, it will highlight some differences between field theories and simple mechanical systems (i.e. those with a finite number of degrees of freedom).

In the case of a system with a finite number of degrees of freedom  the quantum configuration space is just the same as the classical configuration space ${\mathcal{C}}$. The Hilbert space of the system is taken to be $L^2({\mathcal{C}},d\mu)$ where $d\mu$ is an appropriate measure. A simple example would be a particle moving in a Coulomb potential (a good model for the hydrogen atom). The classical and quantum configuration space is ${\mathbf{R}^3}$ and the Hilbert space for the system is $L^2({\mathbf{R}}^3,d\mu)$ where $d\mu$ denotes the standard Lebesgue measure in $\mathbf{R}^3$.

What would be the appropriate choice for a field theory? Let us consider, for example a real scalar field $\phi$ defined on $\mathbf{R}^3$. Even though we make no hypotheses here about its dynamics we will suppose that it is compatible with special relativity (so that we will end up having a description in $\mathbf{R}^4$ endowed with a Minkowskian metric). The classical configuration space $\mathcal{C}_{KG}$ is selected once we impose some smoothness requirements on $\phi$ (for example $\phi\in\mathcal{C}^\infty_0({\mathbf{R}^3})$, that is, smooth functions of compact support defined on $\mathbf{R}^3$). A first difficulty that we encounter if we try to use $L^2(\mathcal{C}_{KG},\mathrm{d}\mu)$ as the Hilbert space of the quantum theory is that there is no obvious choice of measure in this space. It is important to remember at this point that there are subtleties associated to the definition of measures in infinite dimensional topological vector spaces, for example there are no translation invariant measures. At the end of the day a procedure that proves to be very effective in getting a suitable Hilbert space is that of allowing for a suitable modification of the configuration space. The way to do this (and find the appropriate measure) goes back to Kolmogorov and requires the introduction of the so-called cylindrical functions and measures. The advantage of this method is that it can be adapted to the Ashtekar formulation of general relativity and used to build the right Hilbert space for $SO(3)$ connection theories.


Cylindrical functions $\Psi:\mathcal{C}_{KG}\rightarrow{\mathbf{R}}$ are defined with the help of \textit{probes} that are designed to extract partial information from elements of $\mathcal{C}_{KG}$. In this case a convenient choice is given by linear functionals $\displaystyle h_e[\phi]:\mathcal{C}_{KG}\rightarrow\mathbf{R};\phi\mapsto\int_{\mathbf{R}^3}e\phi\mathrm{d}\mu$ defined in terms of a set $\mathcal{S}$ of probe functions $e:\mathbf{R}^3\rightarrow\mathbf{R}$. Given a particular $e\in\mathcal{S}$ some partial information on $\phi$ can be obtained by looking at $h_e$. For example, if $e$ is peaked around a certain point $x_0$, the value of $h_e[\phi]$ gives an approximation for $\phi(x_0)$.

A function $\Psi:\mathcal{C}_{KG}\rightarrow\mathbf{R}$ will be called cylindrical if there exists a finite number of functions $e_1,\ldots,e_n\in\mathcal{S}$ and a smooth $\psi:\mathbf{R}^n\rightarrow\mathbf{R}$ such that for all $\phi\in\mathcal{C}_{KG}$ we have $\Psi(\phi)=\psi(h_{e_1}[\phi],\ldots,h_{e_n}[\phi])$. In such a case we say that $\Psi$ is cylindrical with respect to $e_1,\ldots,e_n\in\mathcal{S}$. They are called cylindrical because they do not depend on all the ``variables $\phi(x)$'' but only on those probed or selected by the specific choice of $e$'s. Cylindrical functions with respect to a fixed set of probes $\alpha=(e_1,\ldots,e_n)$  span a $\mathbf{R}$-vector space that we denote $Cyl_\alpha$. It is easy now to turn $Cyl_\alpha$ into a Hilbert space. By taking a measure $\mu_n$ in $\mathbf{R}^n$
we can define for $\Psi_1,\Psi_2\in Cyl_\alpha$ the following scalar product
\begin{equation}
\langle\Psi_1,\Psi_2\rangle=\int_{\mathbf{R}^n}\bar{\psi}_1\psi_2\,\mathrm{d}\mu_n.
\label{scal_cyl}
\nonumber
\end{equation}
In the previous construction we have been working with a fixed set of probes and we have managed to define a Hilbert space in a rather straightforward way. One of the beautiful features of the previous construction is that it can be extended to the much bigger space of all the cylindrical functions $Cyl:=\cup_{\alpha} Cyl_\alpha$, i.e. the space of functions that are cylindrical with respect to some set of probes. The main problem lies in the construction of an appropriate measure. Although this is not the place to discuss the mathematical details of this construction (the reader is kindly referred to the extensive literature on this subject, and specifically to \cite{AL,ALMMT1} and references therein) it is relatively easy to understand the problems for carrying this program to completion. The main issue is one of compatibility. If we have a function in $Cyl_\alpha$ the scalar product is given by (\ref{scal_cyl}), now, a function in $Cyl$ can be cylindrical with respect to different sets of probes so we must check that if we define our scalar product in $Cyl$ as an extension of the inner product of two such functions this is independent of the set of probes that we use to define it. Notice that we cannot use Lebesgue measures $d\mu_{Leb}$, for example, because $({\mathbf{R}},d\mu_{Leb})$ is not a finite measure space and, then, integrals extended to variables not appearing in the arguments of a certain cylindrical function would give rise to infinite volume factors.

An important result is the fact that the compatibility conditions can be actually be met and it is then possible to define a measure that extends the previous ones to $Cyl$. From the mathematical point of view the actual construction of this measure is carried out by using projective techniques (the final measure is obtained by means of a certain projective limit). The above construction is highly non unique so it is necessary to use some guiding principle to single out a physically interesting measure. An attractive choice in this sense is to use some symmetry if it is available. For example, if we have a Minkowskian background it is natural to exploit the Poincar\'e symmetry to select (or, in fact, narrow down) the measure. Even after doing this there may be some freedom left in the choice of measure because it may depend on physical parameters such as the masses that appear in the Gaussian measures (see \cite{CCQ}). To a large extent the measure is independent of the actual dynamics of the field system but it may depend on some elements necessary to define it, such as the background space-time geometry and the field mass.

Once we have a suitable scalar product $\langle\cdot,\cdot\rangle$ on $Cyl$ we just take the Cauchy completion of $(Cyl,\langle\cdot,\cdot\rangle)$ as the Hilbert space $\mathcal{H}$ of quantum states of the scalar field. An interesting question now is: what do the elements of $\mathcal{H}$ look like? Is it just $L^2({}\mathcal{C}_{KG},d\tilde{\mu})$ for some of the measures mentioned above? If not, can it be thought of as some $L^2$ space defined on an appropriate configuration space? In short the answer to the previous questions is the following: the Hilbert space is $\mathcal{H}=L^2({\mathcal{S}^\prime},d\mu)$ where $\mathcal{S}^\prime$ denotes the topological dual of the space of probes ${\mathcal{S}}$.  State vectors are, then, tempered distributions. The measure $d\tilde{\mu}$ is a regular Borel measure that extends the cylindrical measure defined on $Cyl$.  We have then a situation that is different from the case of a finite number of degrees of freedom because the Hilbert space is now defined on a configuration space that is larger than the classical one. As we can see there is a duality between the probes and the functions in the quantum configuration space. Finally it should be pointed out that the classical configuration space ${\mathcal{C}}_{KG}$ is a zero measure dense subset of ${\mathcal{S}^\prime}$. We hence see that the quantum configuration space is a significant extension of the classical one. In a sense the importance of the classical configurations in the quantum theory is minor in relation with the role played by the distributional elements of the quantum configuration space.

Once we have the Hilbert state of quantum states we must represent the algebra of elementary operators to complete the quantization. In short, configuration (i.e. position) operators act by multiplication
and momentum operators are given by derivations on the ring $Cyl$ that can be interpreted as vector fields in the quantum configuration space $\mathcal{S}^\prime$. The attentive reader may wonder about the relationship between this quantization of the scalar field and the standard Fock one appearing in text book treatments of this problem. The answer is that they are equivalent, in spite of the fact that the procedure that I have sketched here has a real advantage over the standard Fock space construction: it can be generalized to connection field theories. This will be the subject of the next subsection.

\subsection{Connection theories}

The idea now is to follow the steps of the previous construction whenever possible to deal with $SO(3)$ (or rather $SU(2)$) connection field theories. Let us suppose then that we have a theory of $SU(2)$ connections defined on a closed (compact and without boundary) 3-dimensional spatial manifold $\Sigma$; remember that this is the classical configuration space of general relativity in terms of Ashtekar variables. We want to start by building an appropriate space of cylindrical functions as we did for the scalar field discussed in the previous subsection. In order to do this we introduce a set of probe functions to reduce the problem to one with a finite number of degrees of freedom. As the $SU(2)$ gauge invariance that we have in the model will have to be taken into account (and, from the physical point of view, we want to extract gauge invariant information) it is very convenient to look for gauge invariant probes. A natural choice is provided by the holonomies of the $SU(2)$ connection along curves in $\Sigma$. In fact, if we take a sufficiently large set of curves (or graphs) we should be able to recover all the gauge invariant information contained in a given $SU(2)$ connection field. Before I start giving some details about how the construction is carried out several comments about important concepts related to gauge connections should be made.

First of all it is important to point out that due to the fact that $SU(2)$ bundles on oriented 3-dimensional manifolds are trivial, we can indeed represent connections on this type of bundle as $\mathrm{su}(2)$ valued 1-form fields $A_{aA}^{\,\,\,\,B}$ defined in the fundamental representation. In the following it will be convenient to write them as $A_{aA}^{\,\,\,\,\,\,\,\,B}=A_a^i\tau_{iA}^{\,\,\,\,\,\,\,\,B}$ where the three $\mathrm{su}(2)$ Lie algebra vectors are defined in terms of the Pauli matrices as $\tau_i:=\frac{1}{2i}\sigma_i$ and satisfy the commutation relation $[\tau_i,\tau_j]=\epsilon_{ijk}\tau_k$. Holonomies along curves on $\Sigma$ connecting two points $p$ and $q$ define parallel transport of vectors in the fiber of a vector bundle  at $p$ to vectors in the fiber at $q$. They are defined as with the help of the path order operator ${\mathcal{P}}$ according to
$$
u(t):=\mathcal{P}\exp[-\int_0^tA(\gamma^\prime(s))\mathrm{d}s]u_0:=h_\gamma[A]u_0.
$$
Under the action of local gauge transformations of the connection they transform as $h_\gamma[A^\prime]=g(q)h_\gamma[A]g(p)^{-1}$. An important case corresponds to curves $\gamma$ connecting a point $p$ in $\Sigma$ with itself. In this situation the holonomy is an endomorphism of the fiber over $p$ with the important property that its trace, known as the Wilson loop, $W_\gamma[A]:=\mathrm{Tr} h_\gamma[A]$ is gauge invariant. Holonomies play a fundamental role in gauge theories.

As we will be interested in building appropriate measures in spaces of connections it is natural to expect that invariant measures will play a significant role. As is well known, every compact, Hausdorff, topological group has a unique (up to constant factors) left and right invariant measure $\mu_H$ known as the Haar measure. Such measures may also exist for non-compact topological groups but their existence is not guaranteed. Compact Lie groups are finite measure spaces with the Haar measure. In the following we will use this fact to normalize $\mu_H$ to one. This fact is one of the main reasons to work with real connection variables in spite of the difficulties associated with the complicated form of the scalar constraint. As we will see a completely satisfactory construction of a kinematical Hilbert space can be carried out in this case. This would have been much harder if we insisted on using complex variables (remember the discussion about the Immirzi parameter).

As a first step to understand the quantization of $SU(2)$ connection theories it is convenient to consider the simple example of quantum mechanics defined on $SU(2)$ as a configuration space \cite{AL} (this would correspond to the quantization of a spherical top). In order to do this we take the Hilbert space of square measurable functions on $SU(2)$ with respect to the Haar measure $L^2(SU(2),\mathrm{d}\mu_H)$. We define now a representation of the configuration (position) and momentum operators on this Hilbert space. As usual configuration operators  act by multiplication; given a smooth function $f:SU(2)\rightarrow{\mathbf{C}}$ we define
$$(\hat{f}\cdot\Psi)(g)=f(g)\Psi(g).$$
Momentum operators can be associated to smooth vector fields on $SU(2)$. Given a vector field $X\in\mathfrak{X}(SU(2))$ the corresponding momentum operator is defined in terms of the Lie derivative of the wave function along $X$ and the divergence $\mathrm{div}X$ as
$$
(\hat{P}_X.\Psi)(g)=i[\mathcal{L}_X\Psi+\frac{1}{2}(\mathrm{div} X)\Psi](g).
$$
To each vector $v\in\mathrm{su}(2)$ it is possible to associate two natural left and right invariant vector fields $L_v$ and $R_v$ on $SU(2)$, in particular we can do this for the Lie algebra vectors $\tau_i$. These give us an orthonormal basis (with respect to the Cartan-Killing invariant metric $\eta_{ij}$) in the algebra  and let us define two sets of operators $\hat{L}_i$ and $\hat{R}_i$ with two important properties: any of the $\hat{L}_i$ commutes with any of the $\hat{R}_i$ and ${\mathrm{div}}\hat{L}_i={\mathrm{div}}\hat{R}_i=0$. Another important operator can be defined now in terms of the $\hat{L}_i$, $\hat{R}_i$ and the Cartan-Killing metric: the Laplace-Beltrami operator
$$
\hat{J}^2=\eta_{ij}\hat{L}_i\hat{L}_j=\eta_{ij}\hat{R}_i\hat{R}_j=-\Delta.
$$
If this is interpreted as a Hamiltonian for a physical system having $SU(2)$ as the configuration space it corresponds to a spherical top with a fixed point in space. For a given choice of vector $v$ the operators $\hat{J}^2$, $\hat{L}_v$ and $\hat{R}_i$ commute\footnote{An important comment to be made here is that these operators are defined in the Hilbert space $L^2(SU(2),\mathrm{d}\mu_H)$. Even though some of them are labeled by elements of the Lie algebra, they are not elements of the Lie algebra themselves nor belong to the enveloping algebra. There is then no contradiction with the fact that $SU(2)$ has rank 1 and, as a consequence, there is only one independent Casimir operator (an element in the enveloping algebra that commutes with all the Lie algebra generators).}. This means that we can find an orthonormal basis of simultaneous eigenstates of these given by functions $D^{(j)}_{mm^\prime}:SU(2)\rightarrow\mathbf{C}:g\mapsto D^{(j)}_{mm^\prime}(g)$ with $j\in \frac{1}{2}\mathbf{N}\cup\{0\}$, and $m,m^\prime\in\{-j,-j+1,\ldots,j-1,j\}$ for each $j$. Their orthogonality means that
$$
\langle D^{(j)}_{mm^\prime},D^{(\ell)}_{nn^\prime}\rangle:=
\int_{SU(2)}\overline{D}^{(j)}_{mm^\prime}(g)D^{(\ell)}_{nn^\prime}(g)\mathrm{d}\mu_H=
\delta_{j\ell}\delta_{mn}\delta_{m^\prime n^\prime}.
$$
They are eigenfunctions of (minus) the Laplace-Beltrami operator $-\Delta$ with eigenvalue $j(j+1)$. The numbers $m$ and $m^\prime$ are the eigenvalues of $\hat{L}_v$ and $\hat{R}_v$.
All these results can be understood as a consequence of the Peter-Weyl theorem (a central result in harmonic analysis on groups).

After this detour let us go back to the question of choosing an appropriate set of gauge invariant probes to extract the gauge invariant content of a connection field. To this end we will start by associating cylindrical functions to fixed graphs. We define here a graph $\gamma$ as a finite set of edges (compact, 1-dimensional, analytic, oriented and embedded submanifolds of $\Sigma$) that only intersect at the end points.
\begin{figure}
  \includegraphics[width=.9\textwidth]{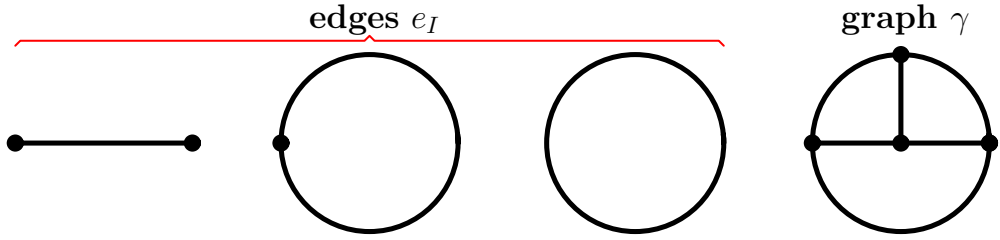}
  \caption{Edges and a graph.}
\end{figure}
Consider now a fixed graph $\gamma$ on $\Sigma$ with $n_\gamma$ edges and $v_\gamma$ vertices and consider the restriction (i.e. the pull back) of any $SU(2)$ connection defined on $\Sigma$ to $\gamma$. For each edge $e_I$ ($I=1,\ldots,n_\gamma$) in $\gamma$ we can get the corresponding holonomy of the connection: $h_{e_I}[A]\in SU(2)$. In this way a $SU(2)$ connection on $\Sigma$ defines a map
$A_\gamma:\gamma\rightarrow[SU(2)]^{n_\gamma}$. By recalling how local gauge transformations act on the holonomies we see that given a graph and a connection, local gauge transformations play a role only at the vertices of the graph. This means that the only gauge freedom that we may have to take care of when restricting ourselves to extracting information from connections by considering holonomies along the edges of a given graph $\gamma$ is tied to the vertices of $\gamma$.

We are ready now to define cylindrical functions associated to graphs (see \cite{AL} and references therein). Let us take as the classical configuration space of our theory the space ${\mathcal{A}}$ of smooth $SU(2)$ connections on $\Sigma$. A function $\Psi:\mathcal{A}\rightarrow\mathbf{C}$ will be called cylindrical if there is a graph $\gamma$ with $n_\gamma$ edges and a function $\psi:[SU(2)]^{n_\gamma}\rightarrow\mathbf{C}$ such that $\Psi(A)=\psi(h_{e_1}[A],\ldots,h_{e_{n_\gamma}}[A])$. This definition implies that a cylindrical function depends on the connection $A$ only through its holonomies along the edges of $\gamma$. As suggested above the holonomies along the edges play the role of ``gauge covariant probes'' to obtain (partial) information about the connections on $\Sigma$. It is easy to give cylindrical functions associated to some graphs. The simplest example would be to take a closed smooth curve without self-intersections and consider the Wilson loop with the holonomies defined in the fundamental representation. A simple and straightforward generalization of this that also gives a cylindrical function is to consider objects of the type $W_\gamma^j[A]\!:=\!Tr[D^{(j)}(h_\gamma(A))]$
where now one takes the trace in the ${\mathrm{D}}^{(j)}$ representation of $SU(2)$. As a final and more general example let us consider the graph shown in figure 3 (where the labels correspond to the representations used for the holonomies along each of the edges)
\begin{figure}
  \includegraphics[width=.2\textwidth]{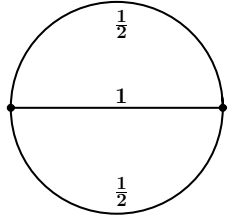}
  \caption{A non-trivial graph used in the example of cylindrical function explained in the text.}
\end{figure}
We have now $D^{(1/2)}_{AB}(h_{e_1}(A))$, $D^{(1)}_{ij}(h_{e_2}(A))$, $D^{(1/2)}_{CD}(h_{e_3}(A))$ where the indices $A,\,B,\,C\,D\in\{-1/2,1/2\}$; $i,j\in\{-1,0,1\}$. In order to get a complex number (remember that cylindrical functions take values in ${\mathbf{C}}$) we can just choose a fixed component for each of the matrices and multiply them (in which case the function would not be invariant under $SU(2)$ transformations on the vertices). If we want to get a gauge invariant object we must contract the indices with a $SU(2)$ invariant object in $\mathrm{D}^{(1/2)}\otimes\mathrm{D}^{(1)}\otimes \mathrm{D}^{(1/2)}$. This is called an intertwiner and can be constructed from symmetrized products of the antisymmetric objects $\epsilon_{AB}$ and $\epsilon^{AB}$. In the present example the (essentially) unique choice is to take the combination of Pauli matrices given by $\sigma_{AC}^i\sigma_{BD}^j$. For vertices of degree 3 these objects can be written in terms of Clebsh-Gordan coefficients or Wigner $3j$ symbols. For higher degrees there are many more choices. It is straightforward to check that a function that is cylindrical with respect to a graph $\gamma$ is cylindrical with respect to any other graph $\gamma^\prime$ that contains it.

As in the scalar case we want now to turn the vector space $Cyl_\gamma$ into a Hilbert space. The role played by Gaussian measures for the scalar field is played now by the Haar measure on $SU(2)$. The fact that $SU(2)$ is compact so that this measure can be normalized is of relevance to ensure the compatibility of the measure that we will introduce (see \cite{Corichi_proc} for a simple example). For two functions $\Psi_1,\Psi_2$ in $Cyl_\gamma$ let us define the scalar product
$$
\langle\Psi_1,\Psi_2\rangle=\int_{[SU(2)]^{n_\gamma}}\prod_{j=1}^{n_\gamma}\mathrm{d}\mu_H(g_j)
\bar{\psi}_1(g_1,\ldots,g_{n_\gamma})\psi_2(g_1,\ldots,g_{n_\gamma})
$$
where $\prod_{j=1}^{n}\mathrm{d}\mu_H(g_j)$ is the Haar measure on $[SU(2)]^{n_\gamma}$. With this scalar product we can define a Hilbert space ${\mathcal{H}}_\gamma$ labeled by the graph $\gamma$. This Hilbert space is essentially the tensor product of several copies of $L^2(SU(2),\mathrm{d}\mu_H)$ associated to each of the edges in $\gamma$. We can now build operators in $\mathcal{H}_\gamma$ from those defined in each $L^2(SU(2),\mathrm{d}\mu_H)$. These are useful to label orthogonal subspaces in a direct sum decomposition of $\mathcal{H}_\gamma$, label the elements of useful orthonormal bases, and also to build the geometric operators associated to areas and volumes. More specifically we can associate operators either to edges or to vertices in the graph. In the case of edges the most interesting operators are obtained by selecting an (oriented) edge $e_I$, one of the vertices $v$ of $e_I$, and a basis $\tau_i$ in $\mathrm{su(2)}$. We can then define $\hat{J}_i^{(v,e_I)}$ as an operator acting on $L^2(SU(2),\mathrm{d}\mu)_{e_I}$ given by $\hat{L}_i$ if $e_I$ starts at $v$ and $\hat{R}_i$ if $e_I$ ends at $v$. With the help of these we can also write $\hat{J}_e^2:=\eta^{ij}\hat{J}_i^{(v,e)}\hat{J}_j^{(v,e)}$ with eigenvalues $j_e(j_e+1)$, $j_e\in\frac{1}{2}\mathbf{N}\cup\{0\}$. The operators defined in this way corresponding to different edges obviously commute.

For vertices the idea is to consider all the edges arriving at or leaving from a fixed vertex $v$ on $\gamma$ and define the operators $\hat{J}^v_i=\sum_{\tilde{e}@ v}\hat{J}_i^{(v,\tilde{e})}$ and $\hat{J}_v^2=\eta_{ij}\hat{J}^v_i\hat{J}^v_j$ where the sum extends to all the edges arriving or leaving from $v$. The $\hat{J}_v^2$ operators have eigenvalues $j_v(j_v+1)$ and commute with the $\hat{J}_e^2$. Given the Hilbert space associated to a certain graph $\mathcal{H}_\gamma$ we can use these operators to split it in a convenient way as $\mathcal{H}_\gamma=\oplus_{j_{e_{I}},j_{v_{\ell}}}\mathcal{H}_{\gamma,j_{e_{I}},j_{v_{\ell}}}$. A comment to be made here is that we can build other operators associated to vertices by using in their definition only a subset of the edges arriving at each of them. These can be used to further decompose the subspaces $\mathcal{H}_{\gamma,j_{e_{I}},j_{v_{\ell}}}$.

At this point we can finally consider the definition of the space of cylindrical functions for $SU(2)$ connection field theories. As we did before we consider now the space of all the cylindrical functions (i.e. cylindrical functions with respect to any graph) $Cyl=\cup_\gamma Cyl_\gamma$. This is now a very large space. The main problem at this point is to define the scalar product for any pair of cylindrical functions (associated to possibly different graphs $\gamma_1$, $\gamma_2$). To this end we first introduce a third graph $\gamma_3$ such that $\gamma_1\subset\gamma_3$ and $\gamma_2\subset\gamma_3$. By doing this we guarantee that both cylindrical functions are cylindrical with respect to the same graph and, hence, can use the definition of scalar product in $\mathcal{H}_{\gamma_3}$ to compute their product. Of course in order to use this definition to define a proper scalar product we need to show that the result is independent of our choice for $\gamma_3$. An important result is that this is indeed the case as a consequence of the left and right invariance of the Haar measure and the fact that we are using normalized measures. In particular the consistency conditions relevant in this case (remember the situation that we encountered in the case of the scalar field) can be met (see \cite{AL} and references therein). An interesting consequence of this definition is the fact that the scalar product of cylindrical functions associated to different graphs (such that they do not have any edges in common) is automatically zero. The last step of the construction of the kinematical Hilbert space $\mathcal{H}$ is to take the Cauchy completion of $Cyl$ with respect to the norm defined by the scalar product.

It is interesting at this point to give a characterization of the elements of $\overline{Cyl}$ that is similar in spirit to the one that we gave for the scalar field. We must then ask ourselves about the possibility of thinking of the Hilbert space that we have built as some $L^2$ defined on some quantum configuration space --that we will refer to as $\overline{\mathcal{A}}$-- with the help of an appropriate measure $d\mu_{AL}$. The quantum configuration space can be characterized in several alternative and interesting ways. Probably the easiest characterization is to describe its elements as quantum connections that assign to each edge $e$ in the spatial manifold $\Sigma$ a $SU(2)$ element in such a way that the two conditions $\bar{A}(e_2\circ e_1)=\bar{A}(e_2)\bar{A}(e_1)$,
$\bar{A}(e^{-1})=(\bar{A})^{-1}(e)$ are satisfied. Notice that these are the only conditions and, in particular, there are no continuity requirements. Another beautiful characterization is the following. The quantum configuration space for $SU(2)$ connection field theories can be understood by considering the Abelian algebra of cylindrical functions and extending it to an abelian $C^*$-algebra with unit $\overline{Cyl}$. This, in turn, can be represented as the space of continuous functions over a compact, Haussdorf space called the spectrum of the algebra $sp(\overline{Cyl}$). We have now $\bar{\mathcal{A}}=sp(\overline{Cyl})$.

What about the measure? The present situation is much more involved that the scalar field case because we are dealing now with an intrinsically non-linear field theory. The compatibility issues that arise here are difficult to deal with but can be successfully solved by resorting again to projective techniques. The family of induced Haar measures that we have introduced for each graph defines a regular Borel measure on $\bar{\mathcal{A}}$ which is invariant under the natural action of diffomeorfisms on $\Sigma$. This is known as the Ashtekar-Lewandowski measure $\mu_{AL}$ \cite{mal} and plays a fundamental role in the definition and description of the kinematical Hilbert space of loop quantum gravity. The fact that such a measure exists is a remarkable --and somewhat unexpected-- result. To put it in a proper perspective it is good to remember that no translation invariant measures exist in topological vector spaces. Several comments are in order now. The first is that it is possible to construct other diff-invariant measures (associated, for example, to knot invariants - see \cite{mal,baez}). The fact that singles out $\mu_{AL}$ among the others is related to the beautiful and far-reaching uniqueness results contained in the LOST and Fleishack theorems \cite{LOST,F} that I will comment on below after the introduction of the flux-holonomy algebra. A second comment concerns the relation between the classical configuration space  ${\mathcal{A}}$ for $SU(2)$ connection theories of the types considered here and the quantum configuration space ${\bar{\mathcal{A}}}$. As shown by Marolf and Mour\~ao \cite{MM} ${\mathcal{A}}$ is densely contained in ${\bar{\mathcal{A}}}$ (in the so-called Gel'fand topology) but has zero measure with respect to $\mu_{AL}$. In plain words, most of the relevant quantum configurations for the connection field are not smooth. A final comment is that the Hilbert space that we have just introduced carries a natural representation of the group of $SU(2)$ gauge transformations and diffeomorphisms (with some technical qualifications related to smoothness conditions for diffeomorphisms). The scalar product is invariant under these transformations and the representation is unitary.

For practical applications, and in particular to describe the action of the geometric operators representing areas and volumes that will be discussed below, it is very useful to introduce a particular orthonormal basis in the Hilbert space ${\mathcal{H}}$ whose construction has been sketched in the previous paragraphs. This is the so-called spin network basis. In principle one could think of writing $\mathcal{H}=\oplus_\gamma\mathcal{H}_\gamma$, but this is not possible because the Hilbert space $\mathcal{H}_{\gamma_2}$ is a non-trivial subspace of $\mathcal{H}_{\gamma_1}$ if $\gamma_2\subset\gamma_1$. This problem can be easily circumvented by considering $\mathcal{H}^\prime_\gamma$ instead of $\mathcal{H}_\gamma$ where the former is defined as  the subspace of $\mathcal{H}_\gamma$ orthogonal to the subspace $\mathcal{H}^\prime_{\tilde{\gamma}}$ associated to every $\tilde{\gamma}$ strictly contained in $\gamma$. Though this definition may sound a little convoluted there is a simple and practical characterization \cite{AL} of this space in terms of the eigenvalues of the operators $\hat{J}_e^2$ associated to the edges of $\gamma$ and the $\hat{J}_v^2$ associated to any spurious vertices that may be present (vertices that do nothing but ``split an edge at a point where $\gamma$ is analytic''). $\mathcal{H}_\gamma^\prime$ is the subspace of $\mathcal{H}_\gamma$ spanned by the simultaneous eigenvectors of $[\hat{j}^v]^2$ and $[\hat{j}^e]^2$ such that the eigenvalues of $[\hat{j}^e]^2$ are not zero for each edge in the graph $\gamma$ and the eigenvalues of $[\hat{j}^v]^2$ are not zero at the spurious vertices.
For a fixed graph $\gamma$ the Peter-Weyl theorem tells us that the product of the functions $D^{(j_1)}_{m_1n_1}(g_1)\cdots D^{(j_{n_{\gamma}})}_{m_{n_\gamma}n_{n_\gamma}}(g_{n_\gamma})$, $j_1,\ldots,j_{n_\gamma}\in\frac{1}{2}{\mathbf{N}\cup\{0\}},\, m_i,\,n_i\in{j_i+\mathbf{Z}}$ with $-j_i\leq m_i, n_i\leq j_i$ are the elements of an orthonormal basis of $\mathcal{H}_\gamma$. This basis can be used in a straightforward way to construct an orthonormal basis for $\mathcal{H}^\prime_{\tilde{\gamma}}$ just by taking $j_1,\ldots,j_{n_\gamma}\neq 0$. The orthonormal basis in the full Hilbert space $\mathcal{H}$ would then be:
$$
\bigcup_{\gamma}\Big\{D^{(j_1)}_{m_1n_1}\otimes\cdots\otimes D^{(j_{n_{\gamma}})}_{m_{n_\gamma}n_{n_\gamma}}:j_i\in \frac{1}{2}\mathbf{N},\,m_i,\,n_i\in j_i+
\mathbf{Z}, \,-j_i\leq m_i, n_i\leq j_i\Big\}.$$
The theory of angular momentum in angular mechanics can be used to look for other bases associated to other commuting sets of angular momentum operators of the type described before.

\section{Quantum geometry}


The purpose of this section is to give an introduction to quantum geometry. To this end it is necessary to first discuss the definition and introduction of the elementary quantum operators that are the building blocks used to define the all-important area and volume operators \cite{AL,Area,Volume}. I will start by introducing the algebra of elementary classical operators \cite{ACZ}. As discussed in the previous section, the classical configuration space ${\mathcal{A}}$ that we are interested in consists of smooth $SU(2)$ connections on $\Sigma$ and the phase space is the cotangent bundle $T^*({\mathcal{A}})$. The information furnished by the symplectic structure in this phase space can be alternatively described by the Poisson brackets between configuration and momentum variables. If we take the smoothed variables
$$^3A[v]:=\int_\Sigma A_a^i\tilde{v}^a_i,\quad \tilde{v}^a_i:\Sigma\rightarrow \mathrm{su}(2)^*,\quad ^3E[f]:=\int_\Sigma \tilde{E}^a_if_a^i,\quad f_a^i:\Sigma\rightarrow  \mathrm{su}(2)$$
we have\footnote{It is customary to write these expressions in distributional form. Here we will have to pay special attention to issues related to the meaning of some smearings of the basic fields so it is important to keep track explicitly of the type of smoothing that we are using. In the following I will drop the prefix $\gamma$ from the expressions for the connection and the densitized triad.}
$$\{^3A[v],^3A[v^\prime]\}=\{^3E[f],^3E[f^\prime]\}=0,\quad \{^3A[v],^3E[f]\}=\int_\Sigma \tilde{v}^a_i f_a^i.$$
This Poisson algebra is not suitable for quantization for a number of reasons. The most important one is the fact that these variables are not gauge covariant in the present non-abelian context. This makes it very difficult to extract physical information (that must be gauge invariant) so one really has to consider other choices. Another important problem is the impossibility of building cylindrical functions with them by exactly following the steps of the previous constructions. The difficulty is that they are not integrable with respect to known diff-invariant measures. For these reasons one must try to find other variables. A natural way to do this is to consider other types of smearings, and specifically, distributional ones. This is suggested by the fact that objects such as holonomies of the $SU(2)$ connection (one-dimensional smearings) \textit{are} gauge invariant. There are other important requirements that a choice of basic configuration and momentum variables must satisfy, in particular they must separate phase space points and their Poisson brackets must define a Lie algebra. With the previous comments in mind it is quite straightforward to put forward a natural set of configuration variables: take cylindrical functions of the connection as defined in the previous section.

The choice of momentum variables is slightly subtler: instead of considering covectors it is useful to take advantage of the fact that momenta can be naturally identified with vector fields on the configuration space. The Hamiltonian vector fields for the 3-dimensional smearings of the triads introduced above are well defined and, hence, we can compute $\{\Psi_\alpha,^3E[f]\}$ ($\alpha$ denotes a graph and $\Psi_\alpha$ is a cylindrical function), however \textit{this Poisson bracket does not give a cylindrical function}. We are then forced to make another choice. To this end let us consider instead a 2-dimensional smearing\footnote{This is the reason why the resulting variables will be called \textit{fluxes}.}. This can be defined rigorously by taking certain limits of the 3-dimensional smearings considered before
$$
\lim_{\epsilon\rightarrow0}\{\Psi_\alpha,{^3}E[{^\epsilon}f]\}.
$$
Here we are using a family of smearing functions ${^\epsilon}f_a^i$ depending on a real parameter $\epsilon$ and such that in the limit $\epsilon\rightarrow 0$ they tend to a distribution with support on a surface [${^\epsilon}f_a^i(x,y,z)=h_\epsilon(z)(\nabla_az)f^i(x,y)$]. The Poisson bracket between a cylindrical function and one of these new smeared variables can be obtained as the limit
\begin{equation}
\displaystyle\lim_{\epsilon\rightarrow0}\{\Psi_\alpha,{^3}E[{^\epsilon}f]\}=\frac{1}{2}
\sum_p\sum_{I_p}\kappa(I_p)f^i(p)X_{I_p}.\psi
\label{eq.1}
\end{equation}
where $p$ are the intersection points of the graph and the surface, $e_{I_p}$ are the edges at each intersection point $I_p=1,\ldots,n_p$, and $\kappa(I_p)$ is $+1$ if the edge lies completely ``above'' $S$, $-1$ if it is ``below'' $S$ and 0 if it is tangent ($S$ is taken to be oriented). Finally $X_{I_p}.\psi$ is the action of the $i$-th left (respectively right) invariant vector field on the $I_p$-th  argument of the function $\psi:[SU(2)]^N\rightarrow\mathbf{C}$ if $e_{I_p}$ points away from (respectively towards) $S$. It is important to note here that with this new type of smearing the Poisson bracket gives a cylindrical function so we are moving in the right direction. In fact, if we manage to interpret the previous limit as the Poisson bracket of $\Psi_\alpha$ with something this would be a good candidate for a momentum variable. The previous choice of ${^\epsilon}f_a^i$ suggests that this variable will be the flux $E[S,f]$ associated to a certain $f^i$ and a choice of a surface $S$.
This is indeed the case when some conditions on the $f^i$'s, and the surface $S$ are imposed. The first is that the functions $f^i$ must be continuous; also the surface $S$ must have the form $\bar{S}-\partial \bar{S}$ with $\bar{S}$ a compact analytic oriented 2-dim submanifold of $\Sigma$. In particular it must have no boundary. The analyticity condition (remember that we are working with piecewise analytic curves to define the holonomies) is used to avoid the appearance of infinite (non-trivial) intersections of $S$ with the edges of the graph $\alpha$. At this point we know the Poisson brackets between configuration variables (that trivially commute) and configuration and momentum variables (given by (\ref{eq.1})). What about the Poisson bracket between two momentum (flux) variables. One would naively expect $\{E[S,f],E[S^\prime,f^\prime]\}=0$ because ``momenta commute''. However this cannot be true for the 2-dimensional smeared variables, the reason: this is incompatible with the expression that we have found for $\{\Psi_\alpha,E[S,f]\}$ because the Jacobi identity would be violated. This can be seen \cite{ACZ} by computing for a simple cylindrical function (a Wilson loop $W_\alpha$)
$$
\{\{E[S,f],E[S,g]\},W_\alpha\}+\{\{W_\alpha,E[S,f]\},E[S,g]\}+
\{\{E[S,g],W_\alpha\},E[S,f]\}
$$
The last two terms can be explicitly computed \cite{ACZ} with the help of (\ref{eq.1}) and, generically, are not zero. Hence the first term cannot be zero if we want the Jacobi identity to be satisfied.

Let us try to understand how the distributional smearings that are being used change the naive expectation about commutativity of the momentum variables. As commented above, for classical finite dimensional systems there is a natural isomorphism between the space of momentum variables and a space of suitably regular vector fields on the configuration space. Momenta can be thought of as vector fields in the configuration space when the phase space is a cotangent bundle (a fact that is translated into the action of the momenta operators after the quantization). Given a fixed vector field $v^a$ on the configuration space we can write variables of the type $P(v)(q,p)=v^a(q)p_a$. Now if $f,f^\prime$ denotes suitable regular functions on the configuration space $\mathcal{C}$ and $v,v^\prime$ are regular vector fields we have
$$\hspace{-5mm}\{Q(f),Q(f^\prime)\}=0,\, \{Q(f),P(v)\}=Q(\mathcal{L}_vf),\, \{P(v),P(v^\prime)\}=-P(\mathcal{L}_vv^\prime).
$$
The last expression tells us that whenever two vector fields $v$ and $v^\prime$ do not commute the Poisson bracket of the associated momenta does not commute either. Going back to our system we can mimic the previous procedure and associate a vector field $X_{S,f}$ to the flux variables $E[S,f]$ defined with the help of a surface $S$. We do this by taking (see (\ref{eq.1}))
 $$
X_{S,f}\cdot\Psi_\alpha:=\frac{1}{2}\sum_p\sum_{I_p}\kappa(I_p)f^i(p)X_{I_p}\cdot\psi.
$$
Some comments are in order now. The first is that $X_{S,f}$ is a vector field in the sense that it is a derivation on the ring $Cyl$. As the commutator of two derivations is a derivation and they form a Lie algebra  there is no problem now with the Jacobi identity. Second, only those derivations that can be obtained by taking finite linear combinations and a finite number of brackets are considered here.
It is possible to explicitly check now that the vector fields $X_{S,f}$ do not commute on $Cyl$. In fact \cite{ACZ}:
$$
[X_{S,f},X_{S^\prime,f^\prime}](\Psi_\alpha)=\frac{1}{4}\sum_p f^i(\bar{p})f^{\prime i}(\bar{p})\epsilon_{ijk}\bigg(
\sum_{I_{p}^{uu^{\prime}}}X^k_{I_{p}^{uu^{\prime}}}-
\sum_{I_{p}^{ud^{\prime}}}X^k_{I_{p}^{ud^{\prime}}}-
\sum_{I_{p}^{du^{\prime}}}X^k_{I_{p}^{du^{\prime}}}-
\sum_{I_{p}^{dd^{\prime}}}X^k_{I_{p}^{dd^{\prime}}}
\bigg)(\psi)
$$
where the sums extend to the vertices of the graph $\alpha$ defining the cylindrical function $\Psi$ that belong to the intersection of the surfaces $S$ and $S^\prime$ (the labels $u$, $u^\prime$, $d$, and $d^\prime$ are used to denote edges that lie above or below $S$ or $S^\prime$ respectively). It is clear now that the two vector fields do not commute generically and then the issue of the non-commutativity of the momentum variables is solved. The underlying reason is related to the way the 2-dim smeared things are obtained as limits from the 3-dim ones. If we denote $^3X[f]$ the vector fields associated with $^3E[f]$ we have
\vspace*{-2mm}
$$
[X_{S_2,f_2},X_{S_1,f_1}].\Psi_\alpha=
\lim_{\epsilon_2\rightarrow0} {^3X}[{^{\epsilon_2}}f]
(\lim_{\epsilon_1\rightarrow0} {^3X}[{^{\epsilon_1}}f]\cdot\Psi_\alpha)-
\lim_{\epsilon_1\rightarrow0} {^3X}[{^{\epsilon_1}}f]
(\lim_{\epsilon_2\rightarrow0} {^3X}[{^{\epsilon_2}}f]\cdot\Psi_\alpha).
$$
It can be seen that the action of vector fields and taking limits does not commute when acting on cylindrical functions and, hence, the previous expression is not zero. It is straightforward to see that generic derivations need not correspond to any phase space function. For example, the commutator $[X_{S_1,f_1},X_{S_2,f_2}]$ with both surfaces intersecting on a curve is a derivation on $Cyl$ but its action is only non-trivial on graphs with edges passing through $S_1\cap S_2$. The commutator has now 1-dim support and then is not a linear combination of fluxes $E[S,f]$. The set of all derivations would be too large to consider here in a definite sense (one would need to incorporate some extra conditions --anticommutation relations-- that introduce additional complications in the formalism). The set formed by the $X_{S_1,f_1}$ is sufficiently small and avoids these difficulties. An important consistency requirement that the variables introduced above indeed satisfy is that they suffice to separate points in phase space. A final comment that will prove to be specially relevant when we consider the quantization of these variables is the fact that they are in a sense hybrid. Whereas the momentum variables --the fluxes-- are linear in the triad field, the configuration variables, obtained from holonomies of the connection, are non-linear (exponential). When we go to the quantum theory the representation that we will be using is something like using $q$ and $\exp(i\beta p)$ as the elementary variables for the quantum particle.

Quantization is straightforward, configuration variables represented by complex valued cylindrical functions on $\mathcal{\bar{A}}$ act by multiplication on the wave functions
$$
(\hat{f}\Psi)[\bar{A}]=f(\bar{A})\Psi[\bar{A}]
$$
and the action of momentum operators (fluxes) is given by
$$
(\hat{E}[S,f]\Psi)[\bar{A}]=i\{E[S,f],\Psi\}[\bar{A}]=
\frac{1}{2}\sum_{v}f^i(v)\sum_{e@v}\kappa(e)\hat{J}_i^{(v,e)}\Psi[\bar{A}].
$$
The sum is carried out over the vertices of $\alpha$ where it intersects $S$. This operator is essentially self adjoint in its domain and, hence, admits a unique extension to the full $\mathcal{H}$. Commutators are represented as $i\hbar$ times the classical Lie bracket between the corresponding classical variables. There are no anomalies in the quantization; commutators exactly mimic the Poisson algebra between classical elementary variables (known as the holonomy algebra or ACZ algebra \cite{ACZ}).

This is the point to state the important uniqueness result mentioned above i.e. the fact there exists exactly one Yang-Mills gauge invariant and diffeomorphism invariant state on the quantum holonomy-flux $\ast$-algebra. This is known as the LOST theorem after their authors (Lewandowski, Okolow, Salhmann, and Thiemann \cite{LOST}) and was also found in a slightly different form by Fleischhack \cite{F}. The key importance of this result is that it shows that the Ashtekar-Lewandowski measure $\mu_{AL}$ is the only diff-invariant measure that supports the representation of the holonomy-flux algebra.

\subsection{Geometric operators}

This last part of the paper will be devoted to a discussion of the most important geometric operators. The most important ones are the length operators, the area operator and the volume operator (\cite{Area,Volume}). Among them the last two are the more interesting in concrete applications. The area operator is important because it shows up in the study of black hole entropy and the volume operator plays a central role in the proper definition of the quantized version of the all-important Hamiltonian constraint. These operators can be rigorously defined in the Hilbert space used above and their spectra can be studied with the help of the spin network basis mentioned above, that is specially well adapted to their description. As we will see these spectra are discrete and are multiples of the Planck area and volume.

Let us start by studying the area operator. To this end let us take a closed (two-dimensional) surface embedded in $S\subset\Sigma$. The densitized triad $\tilde{E}^a_i$ encodes the metric information, hence we can write the area of a surface in terms of it. If we use coordinates on the surface $\sigma_1$ and $\sigma_2$ and then choose a normal $n^a$ to the points of $S$ the area, as a function(al) of $\tilde{E}^a_i$ takes the form
$$
A_S[\tilde{E}^a_i]=\int_S(\tilde{E}^a_i\tilde{E}^b_j\delta^{ij}n_an_b)^{1/2}.$$
We want now to quantize the operator $A_S[\tilde{E}^a_i]$. This means that we have to define its action on the vectors in $\mathcal{H}$. To this end we need to know how it acts on the elements of the orthonormal basis that we have introduced above. A reasonable way to approach this problem is trying to express it in terms of the flux operators $E[S,f]$. The idea is to decompose $S$ in $N$ two dimensional cells $S_I$ of ``small coordinate'' size, use the three Lie algebra vectors $\tau_i$ as test fields $f^i$, and use the flux variables $E[S_I,\tau^i]$ on each cell. We consider then $$A_N[S]:=\gamma\sum_{l=1}^N(E[S_I,\tau_i]E[S_I,\tau_j]\eta^{ij})^{1/2}.$$
This is an approximate expression for the area (``Riemann sum'') in the sense that if the number of cells goes to infinity in such a way that their coordinate size uniformly goes to zero we recover the area in the limit $N\rightarrow\infty$. In order to quantize it we take advantage of the fact that in each cell $E[S_I,\tau_i]E[S_I,\tau_j]\eta^{ij}$ is a positive self adjoint operator on $\mathcal{H}$ (it then has a well defined square root). The action of this operator on an element of $\mathcal{H}_\alpha$ for a fixed graph is straightforward to obtain. The idea is to refine the partition so that every elementary cell has, at most, one transverse intersection with the graph. In this case the only terms contributing come from the $S_I$ that intersect $\alpha$. Once this point is reached further refinements do nothing. The resulting operator can be written as the following sum over the vertices of $\alpha$ that lie on $S$
$$
\hat{A}_{S,\alpha}=4\pi\gamma\ell_P^2\sum_v(-\triangle_{S,v,\alpha})^{1/2}
$$
where $\ell_P$ denotes the Planck length and  $\triangle_{S,v,\alpha}$ is an operator defined on $Cyl$ given by a quadratic combination of the $\hat{L}_i$ and $\hat{R}_i$ operators associated to each edge leaving or arriving at the $v$'s appearing in the previous sum \cite{AL}.

The previous expression is defined on $\mathcal{H}_\alpha$ for a fixed graph. In order to see if it is defined on the whole Hilbert space $\mathcal{H}$ one has to check some consistency requirements related to the fact that a function may be cylindrical with respect to different graphs. It is possible to prove that this is always possible. This operator can be extended as a self-adjoint operator to the full Hilbert space $\mathcal{H}$. It satisfies the important property of being $SU(2)$ invariant and diff-covariant.  The eigenvalues of the area operator are given in general by finite sums of the form
$$
4\pi\gamma\ell_P^2\sum_{\alpha\cap S}[2j^{(u)}(j^{(u)}+1)+2j^{(d)}(j^{(d)}+1)-j^{(u+d)}(j^{(u+d)+1})]^{1/2}
$$
where the $j^{(u)}$, $j^{(d)}$, and $j^{(u+d)}$ are half integers that appear as eigenvalues of the angular momentum operator for the edges that appear in the expression of the area operator (subject to some constraints in the form of inequalities). It is possible to obtain simple expressions for the area operator if we restrict ourselves to the (internal) gauge invariant subspace of $\mathcal{H}$ (this is spanned by the elements of $Cyl$ with vanishing eigenvalues for the vertex operators $\hat{J}^v_i$). For example, if the intersections of $\alpha$ and $S$ are just 2-degree vertices the spectrum of the area operator takes now the simple form
$$
8\pi\gamma\ell^2_P\sum_I(j_I(j_I+1))^{1/2}.
$$
Notice that the Immirzi parameter $\gamma$ appears in all these expressions. This means that it is not an irrelevant arbitrariness in the definition of the canonical transformations leading to the Ashtekar formulation but, rather, a parameter that may show up in (eventually) observable magnitudes such as areas. An interesting pictorial interpretation of the previous expression is the following: The ``quantum excitations of geometry'' are 1-dimensional and carry a flux or area. Any time a graph pierces a surface it creates a quantum of area. This picture is completely different from the Fock one where quantum excitations are interpreted (in fact, behave as) particles. A final comment about the area operator, that highlights its quantum nature is the fact that area operators associated to different intersecting surfaces fail to commute. This implies, in particular, that it is impossible to diagonalize them simultaneously. Although this is conceptually very important, this lack of commutativity has no observable consequences at macroscopic scales.

I will discuss now the volume operator. As I said before it is interesting not only owing to its geometric interpretation but also as an important building block for the construction of the quantized Hamiltonian constraint. The strategy to define it is similar to the one followed for the area operator. The volume of a 3-dim region $B$ (a certain open subset of $\Sigma$) is classically given by
$V_B=\int_B\sqrt{h}$. This can be expressed in terms of the triad as
$$
V_B=\int_B \left|\frac{1}{3!}\epsilon_{abc}\epsilon^{ijk}\tilde{E}^a_i\tilde{E}^b_j\tilde{E}^c_k\right|^{1/2}.
$$
As before we want to rewrite this expression in terms of the basic flux operators. To this end we divide $B$ in cells of a small coordinate volume. In each cell we introduce three surfaces such that each of them splits it in two disjoint pieces.\footnote{This defines the so-called internal regularization. Other (external) regularizations that slightly differ from this one are also found in the literature \cite{AL}. The volume operators obtained in these cases differ from the one discussed here.} An approximate expression for the volume in terms of flux operators is then
$$
\sum_{cells}\left| \frac{(8\pi\gamma\ell_P^2)^3}{6}\epsilon_{ijk}\eta_{abc}E[S^a,\tau^i]E[S^b,\tau^j]
E[S^c,\tau^k]\right|^{1/2}.
$$
When the coordinate size of the cells goes to zero this gives the volume of the region $B$. As in the case of the area operator we define a family of operators for each graph $\alpha$ (satisfying similar consistency conditions) given by
$$
\hat{V}_{B}:=c\sum_v\left|\frac{(8\pi\gamma\ell_P^2)^3}{48}
\sum_{e_1,e_2,e_3}\epsilon_{ijk}\epsilon(e_1,e_2,e_3)\hat{J}_i^{(v,e_1)}
\hat{J}_j^{(v,e_2)}\hat{J}_k^{(v,e_3)}\right|^{1/2}
$$
where $c$ is an undetermined constant and $\epsilon(e_1,e_2,e_3)$ is the orientation factor of the family of edges $(e_1,e_2,e_3)$. Some comments are in order now. First of all it is important to note that the removal of the regulator (i.e. of the auxiliary partition used to define the volume operator) is non-trivial now because the volume operator obtained by ``just taking the limit'' keeps some memory of the details of the partition. Nevertheless there is a way to handle this issue (see \cite{AL}). The orientation factor is zero if the vectors tangent to the edges $e_1$, $e_2$, $e_3$ are linearly dependent at the point where they meet. This means, in particular, that the volume operator is zero when acting on state vectors defined on \textit{planar graphs} i.e. graphs such that at each vertex the tangent vectors are contained in a plane. The volume operator is also zero when acting on gauge invariant states if the vertices are at most of degree 3 (tri-valent vertices). Second, the volume operator has nice invariance properties like the area operator, in particular, it is $SU(2)$ gauge invariant and diffeomorphism covariant like the area operator; the total volume operator (i.e. $\hat{V}_\Sigma$) is diff-invariant. Finally the spectrum of the volume operator is complicated but its features are essentially understood. The eigenvalues are real and discrete; they are not known in general but can be computed in many interesting cases, in particular, when the vertices ar four-valent. The volume operator plays a central role in the implementation of the quantum constraints because the quantum version of the scalar constraint can be written in (relatively) simple terms by using nested Poisson brackets of the total volume operator and the basic canonical variables.

\section{Epilogue}

I hope that the reader of these notes has been able to get a rough idea about the problems faced in the quantization of general relativity and the solutions provided by loop quantum gravity, in particular the implications of diff-invariance in the rigorous and successful quantization of gravity. The issues discussed in the first part of the paper have been focused on the justification of diff-invariance and the rest of the paper has tried to highlight the main points in the construction of the appropriate Hilbert space and some interesting geometric operators. As mentioned in the introduction some very important issues have been set aside. I just want to list them here. First of all it is crucial to understand quantum dynamics i.e. the implementation of the constraints in the quantum theory to determine the Hilbert space of physical states $\mathcal{H}_{phys}$ \textit{including its scalar product!} This is a highly non-trivial step not yet complete at this point. The final success of the whole LQG program probably hinges on this specific point. A related problem has to do with the definition of proper quantum observables, that is, operators that commute with all the constraints. The geometric operators on the kinematical Hilbert space that I have described \textit{do not commute} with the constraints so they are not observables in the Dirac sense. Finding geometrical observables is non-trivial. In fact, there are arguments suggesting that their properties may be quite different from the ones discussed above. There has been a lively debate recently about such issues as the persistence of the discreteness of the spectrum for area and volume operators.

A very important point that must be addressed concerns applications. In this  sense it is remarkable that important physical problems have been already successfully addressed in spite of the fact that the theory is not yet complete. To mention just a few let me call the attention of the reader to the understanding of black hole entropy and the lessons that are being learned these days on cosmology (with the blooming of loop quantum cosmology \cite{ALB}). The results obtained here are very suggestive and give tantalizing glimpses on the physics of the primitive universe and singularities.

I hope that we are fortunate enough to succeed in completing the crucial steps necessary to have a well defined quantum theory of gravity. When that goal is achieved we still face the interesting problem of finding useful ways to extract the relevant physics. This will certainly require a great deal of work and will provide exciting and difficult problems for the future of the field.

\bibliographystyle{aipproc}

\begin{thebibliography}{9}

\bibitem{Wald 1} R. M. Wald, \emph{Quantum Field Theory in Curved Spacetime and Black Hole Thermodynamics}, University of Chicago Press, (1994).

\bibitem{Wein}
S. Weinberg, \emph{The quantum theory of fields. Vol. 2: Modern applications},
Cambridge University Press, (1996).

\bibitem{Barbero:1999ts}
J. F. Barbero G. and  E. J. S. Villase\~nor, Phys. Rev. D61, (2000) 104014.

\bibitem{BarberoG.:2002fd}
J. F. Barbero G. and  E. J. S. Villase\~nor, Phys. Rev. D65, (2002) 125020.

\bibitem{Ashtekar:1986yd} A. Ashtekar, Phys. Rev. Lett. 57, (1986) 2244.

\bibitem{Ashtekar:1987} A. Ashtekar, Phys. Rev. D36, (1987) 1587.

\bibitem{ADM} R. Arnowitt, S. Deser, and C. W. Misner, \textit{The dynamics of general relativity}, in \textit{Gravitation: An Introduction to Current Research,} Ed. L. Witten, Wiley (1962).

\bibitem{Waldbook} R. M. Wald, \textit{General Relativity}, University of Chicago Press, (1984).

\bibitem{Barbero:1994ap} J. F. Barbero G., Phys. Rev. D51, (1995), 5507.

\bibitem{Immirzi} G. Immirzi, Nucl. Phys. Proc. Suppl. 57 (1997), 65

\bibitem{Holst} S. Holst, Phys. Rev. D53, (1996), 5966.

\bibitem{AL} A. Ashtekar and J. Lewandowski, Class. Quant. Grav. 21 (2004) R53

\bibitem{ALMMT1} A. Ashtekar, J. Lewandowski, D. Marolf, J. Mour\~ao, and T. Thiemann, A manifestly gauge invariant approach to quantum gauge theories, in \textit{Geometry of Constrained Dynamical Systems}, Ed. J. M. Charap, Cambridge University Press (1995).

\bibitem{CCQ} A. Corichi, J. Cortez, and H. Quevedo, Annals Phys. 313 (2004) 446.

\bibitem{Corichi_proc} A. Corichi, J. Phys. Conf. Ser. 24 (2005) 1.

\bibitem{mal} A. Ashtekar and J. Lewandowski, Representation theory of analytic holonomy C algebras. In \textit{Knots and Quantum Gravity} ed. John C. Baez, Oxford Lecture Series in Mathematics and its Applications 1 (Oxford University Press, Oxford, 1994).

\bibitem{baez} J. Baez in Proceedings of the Conference on Quantum Topology, ed. David N. Yetter, World Scientific Press, Singapore, (1994) pp. 21.

\bibitem{LOST} J. Lewandowski, A. Okol\'ow, H. Sahlmann, and T. Thiemann, Comm. Math. Phys., Vol. 267 No. 3 (2006), 703.

\bibitem{F} C. Fleischhack, Commun. Math. Phys. 249 (2004) 331.

\bibitem{MM} D. Marolf and J. Mour\~ao, Commun. Math. Phys. 170 (1995) 583.

\bibitem{Area} A. Ashtekar and J. Lewandowski, Class. Quantum Grav. 14, (1997) A55.

\bibitem{Volume} A. Ashtekar and J. Lewandowski, Adv. Theo. Math. Phys. 1, (1997) 388.

\bibitem{ACZ} A. Ashtekar, A. Corichi, and J. A. Zapata, Class. Quantum Grav. 15, (1998) 2955.

\bibitem{ALB} A. Ashtekar, J. Lewandowski, and T. Thiemann, Adv. Theor. Math. Phys. 7 (2003) 233.

\end{thebibliography}

\end{document}